\documentstyle[12pt]{article}
\newcommand{\mysection}{\setcounter{equation}{0}\section}

\begin{document}
\begin {flushright}
EDINBURGH 96/30\\
ITP-SB-96-32\\
LBNL-39025
\end {flushright} 
\vspace{3mm}
\begin{center}
{\Large \bf Resummed heavy quark production cross sections
to next-to-leading logarithm} 
\end{center}
\vspace{2mm}
\begin{center}
N. Kidonakis\footnote{This work was supported in part 
by the PPARC under Grant GR/K54601.} \\
\vspace{2mm}
{\it Department of Physics and Astronomy\\
University of Edinburgh\\
Edinburgh EH9 3JZ, Scotland} \\  
\vspace{2mm}
J. Smith\footnote{This work was supported in part 
under contract NSF 93-09888.} \\
\vspace{2mm}
{\it Deutsches Electronen-Synchrotron, DESY\\
Notkestrasse 85, D-22603, Hamburg, Germany\\
and\\
Institute for Theoretical Physics\\
State University of New York at Stony Brook\\
Stony Brook, NY 11794-3840 }\\  
\vspace{2mm}
R. Vogt\footnote{This work was supported in part by
the Director, Office of Energy Research, Division of Nuclear Physics
of the Office of High Energy and Nuclear Physics of the U. S.
Department of Energy under Contract Number DE-AC03-76SF0098.} \\
\vspace{2mm}
{\it Nuclear Science Division,} \\
{\it Lawrence Berkeley National Laboratory,} \\
{\it  Berkeley, CA 94720} \\
{\it and}\\
{\it Physics Department,}\\
{\it University of California at Davis} \\
{\it Davis, CA 95616}\\
\vspace{2mm}
January 1997
\end{center}
\begin{center} {\bf Abstract}\\
 \begin{quote} \begin{small}
We study how next-to-leading logarithms modify predictions from leading
logarithmic soft-gluon resummation in the heavy quark production 
cross section
near threshold.  Numerical results are presented  for top quark production
at the Fermilab Tevatron and bottom quark  
production at fixed-target energies. 
\end{small} \end{quote} \end{center}

\pagebreak

\mysection{Introduction}
The calculation of hadronic cross sections in the elastic limit
({\it i.e.}, near threshold) in perturbative QCD
involves contributions from the emission of soft gluons. In $n$-th order QCD  
one encounters leading logarithmic (LL) contributions 
proportional to $(-\alpha_s^n/n!) [\ln^{2n-1}((1-z)^{-1})/(1-z)]_+$.
There are also terms with next-to-leading logarithmic (NLL) contributions
proportional to $(-\alpha_s^n/n!) [\ln^{2n-2}((1-z)^{-1})/(1-z)]_+$.
These ``plus'' distributions are large near threshold, $z=1$,
(the precise definition of $z$ will be given in the next section) 
and resummation techniques were originally developed to sum them 
in Drell-Yan production \cite{DY}. As the Drell-Yan process
involves electroweak interactions it has a rather simple color structure. 
Gluons are only radiated from the incoming quark-antiquark pair in the
hadronic collision process so the amplitude is a color singlet. 

The first resummation of LL terms (as well as some NLL terms)
in the heavy (top) quark production cross section 
was discussed \cite{LSN1,LSN2} prior to the discovery of the top 
quark at the Fermilab Tevatron \cite{expt}. The analysis was based on 
the fact that the LL terms are identical to those in the Drell-Yan
process and, even though a cutoff was used to define the 
resummed perturbation series, this paper pointed out the
importance of incorporating resummation effects in threshold
production of heavy quarks in QCD.
Recently other groups have applied more sophisticated 
LL resummation methods to the
top quark production cross section, {\it c.f.}\ Refs.\ \cite{bc,cat}, which
avoid the use of an explicit cutoff, and lead to slightly different 
values for the top quark cross section. At present the experimental
results cannot discriminate between them and there is an ongoing
discussion as to which method is theoretically superior. 
The quark-antiquark annihilation channel is the dominant
partonic process for top quark production in $p \overline p$ collisions 
at center of mass energy $\sqrt{S} = 1.8$ TeV. 
However the gluon-gluon channel is more important in bottom
and charm production near threshold in fixed-target experiments 
with proton and pion beams. 
The LL resummation method of \cite{LSN1} has also been applied to
$b$-quark production at HERA-B \cite{HERAB} and more
generally for fixed-target bottom and charm production by hadron beams 
\cite{sv2}.

The order $\alpha_s^2$ corrections to the Drell-Yan process
in \cite{ZvN} have allowed a check of the NLL terms in the resummation 
formulae, thus the theory is in excellent shape.  However,
this information cannot be used in heavy-quark
resummation since some of the NLL terms are a consequence of 
the more complicated color structure in heavy quark production. 
There is no exact calculation
of the heavy-quark production cross section at order $\alpha_s^4$.  
Even though the NLL terms at order $\alpha_s^3$ for heavy quark production
near threshold were available in 1990 \cite{mssn}, 
at the time of the heavy quark resummation analysis \cite{LSN1}
it was not clear how to resum them. Therefore all work has
concentrated on the LL terms. 
Now this situation has changed. 
A resummation formalism which correctly incorporates
NLL resummation near threshold has recently
been presented by Kidonakis and Sterman \cite{NKGS,Thesis}. 
The authors analytically compared the order-by-order expansion of their 
results with the known NLL terms \cite{mssn}
and found agreement at threshold in both the quark-antiquark channel 
and the gluon-gluon channel.  

In this paper we will apply this new NLL resummation formalism to 
calculate the top quark production cross section at the Fermilab Tevatron 
and the bottom quark production cross section at fixed target energies. 
We are particularly interested in the size and therefore the 
phenomenological importance of the NLL terms with respect to the LL terms.
To determine the relative importance of the NLL terms, we will
use the cutoff method proposed originally in \cite{LSN1}. 
There are several reasons for this, which we now discuss. 
First, as already stated, we are primarily
interested in the corrections the NLL terms 
make to the LL terms. Next, by varying the 
cutoff the failure of the perturbative
expansion and the onset of the nonperturbative
region may be studied directly, bypassed in the newer methods.
Finally, given the complexity of the formulae it is best 
to use the simplest approach to incorporate the NLL terms in a first
test of their magnitude.
An analysis of the NLL terms is very important since 
the present LL resummations are 
based either on neglecting them entirely \cite{bc} or retaining 
only some of them \cite{cat}.
These terms are not universal and could possibly be of importance in
several areas of perturbative QCD besides the analysis
of the heavy quark cross section, including the production of large transverse
energy jets and the production of supersymmetric particles.
The conclusions of our study should not be generalized to these other
reactions since each process is different and requires a separate study.

\mysection{General Formalism}
Here we present some basic formulae which we need for our analysis, based on
the theory developed by Kidonakis and Sterman \cite{{NKGS},Thesis}.
For heavy quark production to be kinematically allowed, the square of the
partonic center of mass energy,
$s = x_a x_b S$, must be larger than the threshold value of the
invariant mass of the heavy quark-antiquark pair, $Q^2 = 4m^2$.  
When the heavy quarks are produced with zero velocity, the true 
threshold and the partonic threshold are equivalent. In either case, 
the plus distributions which must be resummed are functions of
\begin{equation}
z=\frac{Q^2}{s}.
\label{def}
\end{equation}

Our calculations are based on the factorization of
soft gluons from high-energy partons in perturbative QCD
\cite{css,gath}.
All the singular distributions in the heavy quark ($Q$) production 
cross section can be expressed in the form 
\begin{eqnarray}
\frac{d\sigma_{h_1 h_2}}{dQ^2 d\cos\theta^* dy}
&=& \sum_{ab} \sum_{IJ}
\int \, \frac{dx_a}{x_a} \frac{dx_b}{x_b} 
\phi_{a/h_1} (x_a,Q^2) \phi_{b/h_2}(x_b, Q^2)
\nonumber \\
&& \times\delta\Big(y-\frac{1}{2} \ln \frac{x_a}{x_b}\Big)
\hat\sigma^{(IJ)}_{ab}\Big( \frac{Q^2}{x_a x_b S}, \theta^*, \alpha_s(Q^2)
\Big) \,,
\end{eqnarray}
where $x_a$, $x_b$ are the momentum fractions of the
incoming partons, $y$ is the rapidity of the $Q\overline Q$ pair, and
$\theta^*$ is the scattering angle in the pair center of mass system.
The $\phi$'s are parton densities, evaluated at the mass factorization
scale $Q^2$, in the DIS or $\overline{\rm MS}$ mass factorization scheme. 
The hard scattering functions $\hat\sigma$, which depend on the 
color structure  ($I J$) of the
interaction as well as on the partonic channel $a b$ and the
mass factorization scheme, contain all 
the plus distributions in the threshold region.
In \cite{NKGS} it was shown that up to NLL
it is possible to pick a color basis 
in which moments of the functions 
$\hat\sigma^{(IJ)}_{ab}( z, \theta^*, \alpha_s(Q^2))$ exponentiate
with respect to $z$ so that
\begin{eqnarray}
\tilde{\sigma}^{(IJ)}_{ab}(n,\theta^*,Q^2)&=&\int_0^1 dz z^{n-1} 
\hat\sigma^{(IJ)}_{ab}(z,\theta^*,\alpha_s(Q^2))
\nonumber \\ 
&=&h_I (\theta^*,Q^2) h_J^*(\theta^*,Q^2)
e^{E_{IJ}(n,\theta^*,Q^2)} \, .
\label{omegaofn}
\end{eqnarray}
The hard scattering prefactor is a product
of the contributions from the amplitude $h_I$ and its
complex conjugate $h_J^*$. 
Therefore $I$  refers to the decomposition of the
hard scattering amplitude into a color basis, {\it e.g.}\
into an $s$-channel singlet and octet.
The functions $h_I$ and $h_J^*$ have no collinear or soft divergences
at the partonic threshold since these terms have been factored into the 
exponent $E_{IJ}$ in eq.\ (2.3). The exponential in (2.3),
called $S_{IJ}$ in \cite{NKGS,Thesis}, satisfies a renormalization 
group equation with an anomalous dimension matrix $\Gamma_{IJ}$.

The exponents of eq.\ (2.3) are given by
\begin{eqnarray}
&& E_{IJ}^{(ab)}(n,\theta^*,Q^2) = -\int_0^1 dz \frac{z^{n-1}-1}{1-z}
\nonumber \\ &&
\times \Big\{\,(2-r)
\int_0^z \frac{dy}{1-y}\,g_1^{(ab)}[\alpha_s((1-y)^{2-r}(1-z)^r Q^2)]
  +\, g_2^{(ab)}[\alpha_s((1-z) Q^2)]
\nonumber \\ && \qquad \qquad 
+g_3^{(I)}[\alpha_s((1-z)^2 Q^2),\theta^*] 
+g_3^{(J)*}[\alpha_s((1-z)^2 Q^2),\theta^*]\, \Big\}\, ,
\label{Eofn}
\end{eqnarray}
where $g_1$, $g_2$ and $g_3$ are functions
of the running coupling constant $\alpha_s$ with 
$z$- (and $y$)-dependent arguments. 
The parameter $r$, 1 in the DIS scheme and 0 in the $\overline{\rm 
MS}$ scheme, effectively changes the lower limit of 
the $y$-integral in the $\overline{\rm MS}$ scheme ({\it c.f.}\ \cite{cls} and
our later discussion). As it
stands, eq.\ (2.4) is ill-defined since it incorporates arbitrarily soft gluon
radiation at the point 
where the QCD perturbation expansion diverges. There are different 
opinions as to how to define the exponent in a manner 
which clearly separates the
perturbative and non-perturbative regions. We will return 
to this point shortly.

To go to NLL in the exponents $E_{IJ}$,
we need $g_1$ up to and including $O(\alpha_s^2)$
and $g_2$ and $g_3^{(I)}$ to $O(\alpha_s)$.  Refs.\ \cite{bc,cat} 
do not include the NLL
$g_3$ term, treated numerically for the first time here.

The functions $g_1$ and $g_2$ depend on the mass factorization scheme and the
identity of the incoming partons but are essentially independent of the color
structure.  The function $g_1$ is scheme dependent because of the $r$
dependence of the $y$ integration.  It also
depends on the identity of the incoming partons so that
\begin{equation}
g_1^{(ab)}[\alpha_s] = (C_a+C_b)\left ( \frac{\alpha_s}{\pi} 
+\frac{1}{2} K \left(\frac{\alpha_s}{\pi}\right)^2\right )\, ,
\label{g1def}
\end{equation}
where $C_q = C_{\overline q} = C_F$ and $C_g = C_A$. The constant $K$ is
({\it c.f.}\ \cite{kt})
\begin{equation}
K= C_A\; \left ( \frac{67}{18}-\frac{\pi^2}{6}\right ) - 
\frac{5}{9}n_{\rm f}\, ,
\label{Kdef}
\end{equation}
where $n_{\rm f}$ is the number of quark flavors.   
The $g_2$ contribution is only nonzero in the DIS scheme \cite{cls},  
\begin{equation}
g_2^{(q \overline q)\, {\rm DIS}}[\alpha_s]
=-\frac{3}{2}C_F\; \frac{\alpha_s}{\pi}\,, 
\end{equation}
Otherwise, $g_2^{(q \overline q)\, 
\overline{\rm MS}} =
g_2^{gg} = 0$. 

The functions $g_3^{(I)}$ 
depend on the color structure in the hard scattering but are mass
factorization scheme independent. 
In \cite{{NKGS},Thesis} it is shown that
they are the eigenvalues of the anomalous dimension matrix $\Gamma_{IJ}$
which appears in the renormalization group equation for the function $S_{IJ}$.
The precise relation between the $g_3^{(I)}$ and the eigenvalues of
$\Gamma_{IJ}$ is  
\begin{equation}
g_3^{(I)}[\alpha_s((1-z)^2 Q^2),\theta^*]=-\lambda_I[\alpha_s((1-z)^2 Q^2),
\theta^*]\, ,
\end{equation}
showing that $g_3^{(I)}$ and $\lambda_I$ are functions of both the running  
coupling constant $\alpha_s$ (evaluated at a $z$-dependent scale)
and the angle $\theta^*$ between the 
directions of the incoming parton and outgoing heavy quark.
As can be seen below, although the 
eigenvalues are complex, the total $g_3$ contribution to $E_{IJ}^{(ab)}$
in eq.\ (2.4) is real.

Consider first the reaction channel $q(p_a) + \overline{q}(p_b) 
\rightarrow \overline{Q}(p_i) + Q(p_j)$ where $\Gamma_{IJ}$ is two-dimensional.
In the $s$-channel singlet-octet basis the color decomposition
is into $\delta^{ab} \delta^{ij} $ (singlet) and 
$-\delta^{ab}\delta^{ij}/(2N) + \delta^{aj}\delta^{bi}/2$ (octet).
In this basis the components of the matrix $\Gamma_{IJ}$ are \cite{Thesis}
\begin{eqnarray}
\Gamma_{11}&=&-\frac{\alpha_s}{\pi}C_F(L_{\beta}+1+ i\pi ),
\nonumber \\
\Gamma_{21}&=&\frac{2\alpha_s}{\pi}
\ln\left(\frac{u_1}{t_1}\right),
\nonumber \\ 
\Gamma_{12}&=&\frac{\alpha_s}{\pi}
\frac{C_F}{C_A} \ln\left(\frac{u_1}{t_1}\right),
\nonumber \\
\Gamma_{22}&=&\frac{\alpha_s}{\pi}\left\{C_F
\left[4\ln\left(\frac{u_1}{t_1}\right)-L_{\beta}-1- i\pi \right]\right.
\nonumber \\ &&
\left.+\frac{C_A}{2}\left[-3\ln\left(\frac{u_1}{t_1}\right)
-\ln\left(\frac{m^2s}{t_1u_1}\right)+L_{\beta}+ i\pi \right]\right\}\, .
\end{eqnarray}
where
\begin{eqnarray}
L_\beta = \frac{1 - 2m^2/s}{\beta} \left[ \ln \left( \frac{1-\beta}{1+\beta}
\right)  + i \pi  \right] \, \, , 
\end{eqnarray}
and $\beta^2 = 1 - 4m^2 /s$.
The Mandelstam invariants for the reaction are $s=(p_a + p_b)^2$, 
$t = t_1 + m^2 = (p_a - p_i)^2$, and $u= u_1 +m^2 = (p_b - p_i)^2$.

$\Gamma_{IJ}$ is diagonalized in this basis
when the parton-parton c.m. scattering angle 
$\theta^*=90^\circ$ ($u_1=t_1= - s/2$ at threshold) with eigenvalues
\begin{eqnarray}
\lambda_1 & =&  \lambda_{\rm singlet}\,=\,-\frac{\alpha_s}{\pi}
C_F (L_{\beta}+1+ i \pi ) \,, 
\\  
\lambda_2 & = & \lambda_{\rm octet}\,=\,\frac{\alpha_s}{\pi}
\Big\{-C_F (L_{\beta}+1+i \pi )
+\frac{C_A}{2}\Big[L_{\beta}-\ln\Big(\frac{m^2s}{t_1^2}\Big)
+  i\pi \Big]\Big\}\,.
\nonumber
\end{eqnarray}
It is also diagonalized at partonic
threshold $s = 4m^2$ for arbitrary $\theta^*$.

In the partonic channel $g(p_a)+g(p_b) \rightarrow \overline{Q}(p_i)+Q(p_j)$ 
the anomalous dimension matrix is three dimensional. 
In the color basis 
$\delta^{ab}\,\delta_{ji}$\,, $d^{abc}\,\, T^c_{ji}$ and 
$i f^{abc}\,\, T^c_{ji}$\, 
the components of the matrix $\Gamma_{IJ}$ are \cite{Thesis}
\begin{eqnarray}
\Gamma_{11}&=&-\frac{\alpha_s}{\pi}[C_F(L_{\beta}+1)+  C_A i \pi ],
\nonumber \\
\Gamma_{21}&=&0\,,
\nonumber \\
\Gamma_{31}&=&\frac{2\alpha_s}{\pi}\ln\left(\frac{u_1}{t_1}\right),
\nonumber \\
\Gamma_{12}&=&0\,,
\nonumber \\
\Gamma_{22}&=&\frac{\alpha_s}{\pi}\left\{-C_F(L_{\beta}+1)
+\frac{C_A}{2}\left[-\ln\left(\frac{m^2 s}{t_1 u_1}\right)+L_{\beta}-i \pi 
\right]\right\}\,,
\nonumber \\
\Gamma_{32}&=&\frac{N^2-4}{4N}\Gamma_{31}\,,
\nonumber \\
\Gamma_{13}&=&\frac{1}{2}\Gamma_{31}\,,
\nonumber \\
\Gamma_{23}&=&\frac{C_A}{4}\Gamma_{31}\,,
\nonumber \\
\Gamma_{33}&=&\Gamma_{22} \, \, .
\end{eqnarray}

At $\theta^*=90^{\circ}$ the anomalous dimension matrix becomes diagonal 
with eigenvalues
\begin{eqnarray} 
\lambda_1 &=&-\frac{\alpha_s}{\pi}
\left[C_F(L_{\beta}+1)+  C_A i \pi \right]\,,
 \nonumber \\  
\lambda_2 &=&\frac{\alpha_s}{\pi}\Big\{
-C_F(L_{\beta}+1)
+\frac{C_A}{2}\Big[L_{\beta}-\ln\Big(\frac{m^2 s}{t_1^2}\Big)
- i \pi \Big]\Big\}\,,
\nonumber \\ 
\lambda_3 &=&\lambda_2\,.
\end{eqnarray}
We note that $\Gamma_{IJ}$ is also diagonal at partonic threshold 
for arbitrary $\theta^*$.

\mysection{Heavy-quark production cross sections}

As we have already stated, we are primarily interested in the magnitude
of the NLL terms, particularly the $g_3$ contribution,
relative to the LL results of \cite{LSN1,LSN2,HERAB,sv2}.
Therefore we use a similar cutoff scheme, modifying the definitions of 
the exponents in eqs.\ (2.3) and (2.4) to work directly in momentum space and 
avoid working in Mellin space.
This method, introduced in \cite{LSN1},
exploits the correspondence between the
moment variable $n$ and logarithmic terms in the momentum space variable 
$1-z$. To directly compare with other recent LL resummation schemes
in Refs.\ \cite{bc,cat} would, in principle, require three separate studies 
since each group of authors defines eq.\ (2.4) differently. 
Therefore, in this paper we only consider the approach based on \cite{LSN1}.

First, we identify
\begin{eqnarray} 1-z = \frac{s_4}{2m^2} \,,
\end{eqnarray}
where $s_4$ is the invariant mass of the 
heavy-quark + gluon system which recoils against the detected heavy antiquark
in inclusive $\overline Q$ production.
The invariants satisfy
$s + t_1 + u_1 = s_4$ so that in the elastic limit
$s_4 \rightarrow 0$ or $z \rightarrow 1$.  Note that eq.\ (3.1)
differs from the identification $1- z = s_4/m^2$ in Refs.\ \cite{LSN1,bc}. 
The factor of two is necessary to compare 
the NLL terms \cite{Thesis} with the expressions previously given
in \cite{mssn}.  With the identification in eq.\ (3.1)
all the NLL terms in the $q\overline q$ channel are reproduced,
not only exactly at threshold
but also at finite $s_4$ except for one NLL term in $C_F^2$ which 
has a different coefficient. (This small difference is due to the fact that
the $Q\overline Q$ invariant mass and the $Qg$ invariant mass are different
invariants).
A similar observation holds for the $gg$ channel.

Further we exploit the fact that a heavy quark-antiquark
pair will be preferentially produced back-to-back in the parton-parton 
center-of-mass frame near threshold. We also 
choose the case of $\theta^* = 90^\circ$ to use
the results of eqs.\ (2.11) and (2.13). This choice 
avoids both the diagonalization
of the three dimensional matrix in the gluon-gluon channel and 
an extra numerical integration over $\theta^*$.

The resummation is done using eqs.\ (3.17)
and (3.19) of \cite{LSN1} where the ``plus" 
distributions are removed through integration by parts. Therefore
the resummed partonic cross section is defined as
\begin{equation}
\sigma_{ab}(s, m^2)=-\int_{s_{\rm cut}}^{s-2ms^{1/2}} ds_4 
f_{ab}\left(\frac{s_4}{2m^2}\right)
\frac{d \overline{\sigma}_{ab}^{(0)}(s, s_4, m^2)}{ds_4}\, ,
\end{equation}
where $ab = q \overline q$ or $gg$ and $s_{\rm cut}$ will be defined below.
The other inputs needed here are the function $f_{ab}$, 
in either the DIS scheme
or the $\overline{\rm MS}$ scheme and the differential of the
Born cross section.  
The exponential function $f_{ab}(s_4/2m^2)$ is given by a 
momentum space version of eq.\ (2.4).   
This is best illustrated by examining eqs.\ (3.31) and (3.35) 
in \cite{LSN1}. There we showed how to include gluons with
an invariant mass $s_4' > s_4$ and integrate over their
transverse momenta up to the kinematic limit.  The corresponding
result follows by first replacing $z$ in eq.\ (2.4) by $z'$.
Next we replace $z'^{n-1} -1$ in the numerator of eq.\ (2.4), 
by $-1$ and 
introduce the variables $\omega' = 1 -z' = s_4'/(2m^2)$ and
$\xi = (1-y) (1-z') Q^2/\Lambda^2$. By explicitly leaving $Q^2$ in the 
following discussion, we can
later study the scale dependence of the calculation.
Note that $\alpha_s$ in eqs.\ (2.5)-(2.13)  
is a function of $s_4'/2m^2$. 

We first discuss the $q \overline q$ channel.  There is only one kinematic 
structure in this channel so that the singlet and
octet eigenvalues in eq.\ (2.11) multiply the same function.  At this order in
perturbation theory however, only the octet component contributes,
{\it c.f.}\ the discussion in \cite{NKGS}.  Therefore
in the DIS scheme at $\theta^* = 90^\circ$, the exponential function is 
\begin{equation}
f_{q \overline q}^{\rm DIS} \Big( \frac{s_4}{2m^2} \Big) = 
\exp [E_{q\overline q}^{\rm DIS} + E_{q\overline q}(\lambda_2) ]\,,
\end{equation}
where
\begin{eqnarray}
 E_{q\overline q}^{\rm DIS} &=& E_{q \overline q}^{\rm DIS}(g_1) + 
E_{q \overline q}^{\rm DIS}(g_2)
\nonumber \\
&=& \int_{\omega_0}^1\frac{d\omega'}{\omega'}
\Big\{\int_{\omega'^2 Q^2/\Lambda^2}^{\omega' Q^2/\Lambda^2} \frac{d\xi}{\xi}\,
 \Big[ \frac{2 C_F}{\pi} \Big( \alpha_s(\xi) 
+ \frac{1}{2\pi} \alpha^2_s(\xi) K\Big) \Big] 
\nonumber \\ &&  \qquad \qquad 
 - \frac{3}{2} \frac{C_F}{\pi} \alpha_s
\Big( \frac{\omega' Q^2}{\Lambda^2}\Big)
\, \Big\} \,. 
\label{Eofm}
\end{eqnarray}
Since our calculation is not done in moment space, the $\omega'$ integral 
is cut off at $\omega_0= s_4/2m^2$.  Because the running coupling
constant diverges when $\omega'^2 Q^2/\Lambda^2 \sim 1$,
the minimum cutoff in eq.\ (3.2) 
is $s_{\rm cut} = s_{4, {\rm min}} \sim 2 m^2 \Lambda/Q$.  If we choose the
scale $Q^2 = m^2$, $s_{\rm cut} \sim 2m \Lambda$.  
In general we choose a larger value to stay away from the point of divergence.
Once we have chosen a reasonable
cutoff, consistent with the sum of the first few terms
in the perturbative expansion,
we can study the size of the NLL terms with respect to the LL terms
(the order-by-order expansion does not require a cutoff).

We use the two-loop running coupling constant,
\begin{eqnarray}
 && \alpha_s(\xi) = \frac{1}{a \ln \xi}
+ \frac{b}{a} \frac{\ln(\ln \xi)}{\ln^2(\xi)} \,, \nonumber \\
 && a = \frac{11C_A - 4T_f n_{\rm f}}{12\pi} \,,  \nonumber \\
&& b = -6 \frac{17 C_A^2 - (6C_F + 10C_A)T_fn_{\rm f}}
{(11 C_A - 4 T_f n_{\rm f})^2}\,,
\end{eqnarray}
with $C_A = 3$, $C_F = 4/3$ and $T_f = 1/2$. It is obvious that when 
$\xi < 1$, $\ln\xi$ 
is negative so that $\ln(\ln\xi)$ has a cut and needs a precise definition. We 
use the cutoff on the $\omega'$ variable to stop the integration before
that point.  The exact cutoff depends on the quark mass, the scale 
and $\Lambda_{n_{\rm f}}$, the QCD scale parameter for $n_{\rm f}$ light quark
flavors, in the parton densities.  
Both the $g_1$ integrals over
$\omega'$ and $\xi$ and the $g_2$ integral over $\omega'$ as well as the
contributions to the $g_3$ integration which only contain $s_4'$ in the running
coupling constant can be done analytically, as shown in the appendix.
  
In the $\overline{\rm MS}$ scheme,
\begin{equation}
f_{q \overline q}^{\overline {\rm MS}}\Big( \frac{s_4}{2m^2} \Big) = 
\exp [E_{q\overline q}^{\overline{\rm MS}} + E_{q\overline q}(\lambda_2) ]\,,
\end{equation}
where now
\begin{eqnarray}
 E_{q\overline q}^{\overline{\rm MS}} = E_{q \overline q}(g_1)
= \int_{\omega_0}^1\frac{d\omega'}{\omega'}
\int_{\omega'^2 Q^2/\Lambda^2}^{Q^2/\Lambda^2} \frac{d\xi}{\xi}\,
 \Big[ \frac{2 C_F}{\pi} \Big( \alpha_s(\xi) 
+ \frac{1}{2\pi} \alpha^2_s(\xi) K\Big) \Big]\,. 
\label{Eofmsb}
\end{eqnarray}
Note the difference between the upper limits of the $\xi$-integrations 
introduced by the mass factorization scheme in eqs.\ (3.4) and (3.7).

The color-dependent $g_3$ contribution in eq.\ (2.11) leads to 
\begin{eqnarray}
&& \!\!\!\! E_{q\overline q}(\lambda_i)=
-\int_{\omega_0}^1\frac{d\omega'}{\omega'}
 \Big\{
 \lambda_i 
\Big[\alpha_s \Big(\frac{\omega'^2 Q^2}{\Lambda^2}\Big),\theta^*=90^\circ
\Big]  
 + \lambda_i^* 
\Big[\alpha_s \Big(\frac{\omega'^2 Q^2}{\Lambda^2}\Big),\theta^*=90^\circ
\Big]  
\, \Big\} \,.  \nonumber \\ 
\label{Eofni}
\end{eqnarray}
in both mass factorization schemes, where $i= 1,2$.

The differential function in eq.\ (3.2) follows from the kinematic 
behaviour of the Born cross section, represented by
\begin{equation}
F_{q\overline q}^B(s, t_1, u_1) =
\frac{t_1^2 + u_1^2}{s^2} + \frac{2m^2}{s}\, . 
\end{equation}
Substituting 
$t_1 = -1/2 \{ s -s_4 - [ (s-s_4)^2 - 4 s m^2 ]^{1/2} \cos \theta^* \}$
and
$u_1 = -1/2 \{ s -s_4 + [ (s-s_4)^2 - 4 s m^2 ]^{1/2} \cos \theta^* \}$,
and taking $\theta^* = 90^\circ$,  we define, analogous 
to eq.\ (2.20) in \cite{LSN1}
\begin{equation}
\overline F^{(0)}_{q \overline q} = \frac{[ ( s-s_4)^2 - 4sm^2]^{1/2}}{2s^2}
F_{q\overline q}^B \,.
\end{equation}
Differentiating with respect to $s_4$, we find
\begin{equation}
\frac{d \overline{F}^{(0)}_{q \overline{q}}}{ds_4}=-\frac{1}{4s^4}
\frac{s-s_4}{\sqrt{(s-s_4)^2-4sm^2}}[3(s-s_4)^2-4sm^2]\,.
\end{equation}

The integrand in eq.\ (3.2) becomes, for both factorization schemes, 
\begin{eqnarray}
f_{q \overline q}\left(\frac{s_4}{2m^2}\right)
\frac{d \overline{\sigma}_{q \overline{q}}^{(0)}(s, s_4, m^2)}{ds_4}\,
&=& 
\pi \alpha_s^2 K_{q \overline{q}} N C_F \frac{d 
\overline{F}^{(0)}_{q \overline{q}}}{ds_4}
\nonumber \\ && \qquad   
\times \exp [E_{q \overline q}+ E_{q\overline q}(\lambda_2)]
\,,
\end{eqnarray}
where $K_{q \overline{q}}=1/N^2$ is a color average factor.  
We note that near threshold 
$d \overline{\sigma}_{q \overline{q}}^{(0)}/ds_4$ 
is approximately a factor of two smaller than 
$d{\sigma}_{q \overline{q}}^{(0)}/ds_4$ 
in eq.\ (3.20) of Ref.\ \cite{LSN1} where the
angle $\theta^*$ was analytically integrated 
before differentiating with respect to $s_4$.   

The treatment of the gluon-gluon channel is very similar
but now we have three distinct color structures.  However, 
only two of them are independent.  Therefore we define $f_{gg}$ for each
eigenvalue so that $f_{gg,i} = \exp(E_{gg} + E_{gg}(\lambda_i))$ 
with a one-to-one
correspondence between  $E_{gg}(\lambda_i)$ and the eigenvalues 
of eq.\ (2.13) so that
\begin{eqnarray}
&& \!\!\!\! E_{gg}(\lambda_i)=
-\int_{\omega_0}^1\frac{d\omega'}{\omega'}
 \Big\{
 \lambda_i 
\Big[\alpha_s \Big(\frac{\omega'^2 Q^2}{\Lambda^2}\Big),\theta^*=90^\circ
\Big]  
 + \lambda_i^* 
\Big[\alpha_s \Big(\frac{\omega'^2 Q^2}{\Lambda^2}\Big),\theta^*=90^\circ
\Big]  
\, \Big\} \,, \nonumber \\ 
\label{Eofnj}
\end{eqnarray}
where $i= 1$, 2. 
In the $\overline {\rm MS}$ scheme, we take
\begin{equation}
 E_{gg} =
\int_{\omega_0}^1\frac{d\omega'}{\omega'}
 \int_{\omega'^2 Q^2/\Lambda^2}^{Q^2/\Lambda^2} \frac{d\xi}{\xi}\,
 \Big[ \frac{2 C_A}{\pi} \Big( \alpha_s(\xi) 
+ \frac{1}{2\pi} \alpha^2_s(\xi) K \Big) \Big] \,, 
\end{equation}
where $K$ is defined as in eq.\ (2.6). This choice of $K$ 
requires some justification, which we have not found
explicitly in the literature, due to the fact that the discussions
of the corrections to the Drell-Yan process are made in the DIS scheme. 
However, it should be possible to prove rigorously that
(3.14) holds \cite{gs}.

Because there are two kinematical structures, $d\overline 
\sigma^{(0)}_{gg}/ds_4$ has two contributions:
\begin{eqnarray}
F^B_{gg,I}(s,t_1,u_1) &=&  \frac{t_1}{u_1} + \frac{u_1}{t_1}
+ \frac{4m^2 s}{t_1 u_1} \Big( 1- \frac{m^2 s}{t_1 u_1} \Big) \,
\nonumber \\ 
F^B_{gg,II}(s,t_1,u_1) &=  &
\Big( 1 - \frac{4t_1 u_1}{s^2} \Big)F^B_{gg,I}(s,t_1,u_1) \, .
\end{eqnarray}
The second kinematical structure, proportional to the first,
comes from the nonabelian part of the cross section.  As in eq.\ (3.10),
we define
\begin{eqnarray}
\overline F^{(0)}_{gg,i} = \frac{[ (s-s_4)^2 - 4 s m^2]^{1/2} }{2s^2}
F^B_{gg,i} \, \, .  
\end{eqnarray} 

The differential with respect to $s_4$ of the two components yields 
\begin{eqnarray}
\lefteqn{ \frac{d \overline{F}^{(0)}_{gg,I}(s, s_4, m^2)}{ds_4} = 
-\frac{1}{2s^2 \sqrt{(s-s_4)^2-4sm^2}}\Big[2(s-s_4)
-\frac{16m^2s}{s-s_4} } \nonumber \\ && 
+\frac{320m^4s^2}{(s-s_4)^3} -\frac{1024m^6s^3}{(s-s_4)^5}\Big]  \\ 
\lefteqn{ \frac{d \overline{F}^{(0)}_{gg, II}(s, s_4, m^2)}{ds_4} =
\frac{d \overline{F}^{(0)}_{gg,I}(s, s_4, m^2)}{ds_4} } \nonumber \\ && 
-\frac{1}{2s^2\sqrt{(s-s_4)^2-4sm^2}}\left[\frac{512 m^6 s}{(s-s_4)^3}
-\frac{6(s-s_4)^3}{s^2}-\frac{64m^4}{s-s_4}\right]  
\end{eqnarray}
Inserting the remaining color factors yields the $gg$ integrand in eq.\ (3.2). 
\begin{eqnarray}
\lefteqn{ f_{gg}\Big(\frac{s_4}{2m^2} \Big)
\frac{d \overline{\sigma}_{gg}^{(0)}(s, s_4, m^2)}{ds_4} = 
\frac{\pi}{8} K_{gg} N C_F \alpha_s^2 
\Big\{ \Big[2C_F f_{gg,1} \Big(\frac{s_4}{2m^2}\Big) } \\ && 
+ C_F(N^2-4)f_{gg,2}\Big(\frac{s_4}{2m^2}\Big) \Big] 
\frac{d \overline{F}^{(0)}_{gg, I}}{ds_4} 
+ 4C_A f_{gg,2}\Big(\frac{s_4}{2m^2}\Big)
\frac{d \overline{F}^{(0)}_{gg, II}}{ds_4}\Big\} \, \, , \nonumber
\end{eqnarray}
where $K_{gg}=(N^2-1)^{-2}$ is a color average factor. One
can numerically compare the coefficients of the color terms in 
$d\sigma^{(0)}_{gg}/ds_4$ in eq.\ (3.21) of Ref.\ \cite{LSN1}
with the color terms near threshold in $d\overline{\sigma}_{gg}^{(0)}/ds_4$
in eq.\ (3.19). Those in Ref.\ \cite{LSN1} are integrated over the angle
$\theta^*$ before differentiation with respect to $s_4$ and
are several times larger than those found here.  Note that eq.\ (3.19) contains
several terms which can have different signs, making it a more
complicated function of $s_4$. Therefore we do not expect our $gg$ results 
to agree with results obtained without attention to the color decomposition
\cite{LSN1,LSN2,HERAB,sv2}.

Because the $\omega'$ integral is cut off at $s_4/(2m^2)$, the partonic
cross section in eq.\ (3.2)
must also have a lower limit of $s_{\rm cut}$
for the integrals to be finite.  
In earlier work up to LL \cite{LSN1,LSN2,HERAB,sv2},
the lower limit of the partonic cross section, $s_0$, was taken to be
$s_0/m^2 = (\mu_0/\mu)^2$ in the DIS scheme and $s_0/m^2 = (\mu_0/\mu)^3$
in the $\overline{\rm MS}$ scheme where $\Lambda_{n_{\rm f}} \ll \mu_0 \ll 
\mu$ and $\mu$ is the renormalization scale.  It was
assumed that the integral over the nonperturbative region 
below $s_0$ was negligible compared to the integral
over the perturbative region. The final
resummed cross section was obtained by comparison with the order-by-order
approximate cross section up to $O(\alpha_s^4)$ as a function of $\mu_0$. 
Since each term in the 
perturbative result is positive, $\mu_0$ was chosen 
such that the resummed cross section was slightly larger
than the sum of the approximate cross sections. It was found that 
for $\mu = m$ in the DIS scheme
$\mu_0 \approx (0.05-0.1)m$.  Larger values were needed in the 
$\overline{\rm MS}$ scheme because of the cubic power of $\mu_0/\mu$.
In the $q \overline q$ channel, $\mu_0 \approx (0.1-0.2)m$, while $\mu_0 
\approx 0.35m$ in the $gg$ channel.  The $\mu_0$ needed in the $gg$ channel was
larger because of the color factor $C_A$ in $E_{gg}$, see eq.\ (3.14).  
As before, we will use the cutoff method since increasing $s_{\rm cut}$ 
away from $s_{4, {\rm min}}$ reduces the scale and scheme dependence of
the resummed perturbative cross section, as we discuss in detail in the next
section and was also shown previously for the LL resummation \cite{LSN1,LSN2}. 
In the next section, we will calculate the $\mu_0$ appropriate for our
chosen $s_{\rm cut}$.

\mysection{Numerical results}

We first present our results for top quark production.
In fig.\ 1(a) we plot the exponents contributing to $f_{q \overline q}^{\rm
DIS}$ in eq.\ (3.3) at $\theta^* = 90^\circ$
as a function of $s_4/2m^2$ with $m = 175$ GeV/$c^2$ and $\sqrt{s} = 351$ GeV.
Since only the octet component contributes to $g_3$ here, only $E_{q
\overline q}(\lambda_2)$ is shown.  
We also show, for comparison, 
the one-loop calculation of $g_1$, $E^{\rm DIS}$, 
given in \cite{LSN1}.  The one-loop and two-loop results are quite similar
until $s_4$ approaches $s_{4, \rm min}$. As $\omega' Q/\Lambda \rightarrow 
1$, $E_{q \overline q}^{\rm DIS}(g_1)$ is larger since $E^{\rm DIS}$ was 
smoothed near the cutoff \cite{LSN1}.
The sum $E_{q \overline q}^{\rm DIS} + E_{q \overline q}(\lambda_2)$ 
is always larger than $E^{\rm DIS}$ because the $g_3$
contribution compensates for the small, negative $g_2$ component ($|E_{q 
\overline q}^{\rm DIS}(g_2)|$ is shown).  The power of $1-z$ in
$\alpha_s$ determines the slope of the exponents at small $s_4/2m^2$.  
The $g_2$ contribution, linear in $1-z$,
is the flattest as $s_4/2m^2 \rightarrow 0$.  The $g_1$ component shows more
growth, diverging from $E^{\rm DIS}$ near $s_{4, \rm min}$, and the $g_3$
contribution, quadratic in $1-z$, grows fastest.  The  growth of the sum 
is intermediate to that of $E_{q \overline q}^{\rm DIS}(g_1)$ and
$E_{q \overline q}(\lambda_2)$.    

To better illustrate the importance of these contributions to the partonic 
cross section, we show the relative enhancements as a function of $s_4/2m^2$ 
in fig.\ 1(b).  The negative contribution and
the slow growth of $g_2$ near $s_{4, \rm min}$ 
are reflected in the near threshold
behavior of $f_{q \overline q}(g_1 + g_2)/f_{q \overline q}(g_1)$ where {\it
e.g.}\ $f_{q \overline q}(g_1) \equiv \exp(E_{q \overline q}^{\rm DIS}(g_1))$.
In the intermediate range of $s_4/2m^2$,
$E^{\rm DIS}$ is larger than $E_{q \overline q}^{\rm DIS}(g_1)$.
Although $E_{q \overline q}(\lambda_2)$ dominates $f_{q \overline q}^{\rm DIS}$
as $s_4/2m^2 \rightarrow 1$, the enhancement in this region is not very
significant.  Thus the choice of the cutoff determines the
importance of the NLL contributions.  However, a large NLL correction is not
necessarily reflected in the hadronic cross section since it is the parton
luminosity relative to the equivalent $\eta = s/4m^2 - 1$
that determines the true strength of the corrections. 

Similar results are seen for the exponents in the $\overline{\rm MS}$ scheme,
shown in fig.\ 2.  The $g_2$ term vanishes in this scheme.  The change in the 
upper limit of the $\xi$
integral increases the phase space of $E_{q \overline q}^{\overline{\rm 
MS}}$.  Both these changes enhance the $\overline{\rm MS}$ result.   
The $g_3$ contribution is identical to that shown in fig.\ 1.

In fig.\ 3 we plot the corresponding results for the $gg$ channel
in the $\overline{\rm MS}$ scheme.
Note that the real parts of the $\lambda_i$'s are the same in eqs.\ (2.11) and
(2.13) so that the strength of the $g_3$ contributions are the same
in the $q \overline q$ and $gg$ channels.  However, since different color 
structures in the $gg$
channel multiply different derivatives of the Born cross section, 
there is no single sum involving the NLL components as in the $q \overline
q$ channel.  The $E_{gg}(\lambda_1)$ component is quite small and negative
near threshold, making the sum $E_{gg} + E_{gg}(\lambda_1)$ 
indistinguishable from $E_{gg}$.  The $E_{gg}(\lambda_2)$ contribution is
the same as in the $q \overline q$ channel, resulting in the larger sum
$E_{gg} + E_{gg}(\lambda_2)$ shown in the dot-dashed curve. 

Note that changing the scale effectively shifts the range of $s_4/2m^2$.  
We found also that, for the same value of $s_4/2m^2$, the exponents are larger
for smaller values of $Q^2$, as expected.

We have checked the energy dependence of the $g_3$ contributions.  We find that
$E_{gg}(\lambda_1)$ is always negative, decreasing the NLL
correction for all $s$.  Its energy dependence is quite strong and
thus above threshold it serves to damp the resummation.  
On the other hand, there is very little energy
dependence of $E_{IJ}(\lambda_2)$ close to threshold.  As $s$ grows large, the
$\ln(m^2s/t_1^2)$ term in eqs.\ (2.11) and (2.13)
dominates the energy dependence, causing $E_{IJ}(\lambda_2)$ to change sign, 
eventually leading to a strong damping of
the NLL component (and probably leading 
to a smooth energy variation of the resummed cross section above threshold).  
This logarithmic term also
has the only explicit $s_4'$ dependence of the $g_3$ contribution at 
$\theta^* = 90^\circ$ apart from the argument of $\alpha_s$.  
Since $\lambda_1$ only depends on $s_4'$ through $\alpha_s$, these 
contributions have similar behaviours in $s$.  

In fig.\ 4 we plot the partonic $t \overline t$ cross section as a function of
$\eta$ for the $q \overline{q}$ and $gg$ channels.  
At $\eta \approx 1$ the $gg$ contribution is numerically smaller than 
the corresponding $q \overline q$ result.  
The importance of the NLL contributions to the partonic cross section
depends on the lower limit of the $s_4$ integral.  We have chosen $s_{\rm cut}
= 10 s_{4, \rm min}$, corresponding to $\mu_0 = 0.26m$ 
in the $\overline{\rm MS}$
scheme and 0.13$m$ in the DIS scheme when $Q^2 = m^2$, 
similar to the values used previously.
Changing $s_{\rm cut}$ changes the effective $\eta$ range.  However,
note that for $\eta \geq 1$, the $q \overline q$ results are
nearly scheme independent.  The $q \overline q$ result in the $\overline{\rm
MS}$ scheme is always larger than the DIS result.
We have checked that the dependence on $s_{\rm cut}$ is rather weak in this
channel.  The $gg$ channel is similar although the
different color structure of the NLL terms leads to a stronger variation 
over all $\eta$ in this channel. 

To illustrate the contribution of the
NLL terms to the partonic cross section, in fig.\ 4(b) we show the partonic
cross section with only the LL contributions
for the same value of $s_{\rm cut}$.  At $\eta >1$ the $q \overline q$
contributions with and without the NLL components are nearly the same but at
$\eta \leq 0.02$ the $g_3$ contribution makes the partonic cross section
up to a factor of two larger.  Both with and without the NLL terms, the scheme
dependence increases as $\eta$ decreases.

We note that for fixed angle, $\theta^* = 90^\circ$, the partonic cross
section in the $gg$ channel can become negative around $\eta \sim 1-2$.  
Because $d \overline F_{gg, II}^{(0)}/ds_4$ is 
generally opposite in sign to $d \overline F_{gg, I}^{(0)}/ds_4$,
when it grows larger than  $d \overline F_{gg, I}^{(0)}/ds_4$ the partonic $gg$
cross section becomes negative. In our calculation of the hadronic cross
section, we exclude the negative region if it occurs.
For the particular values of $s_{\rm cut}$ and $Q^2$ shown here, the partonic
cross section is always positive.  However, for other values of 
$s_{\rm cut}$ and $Q^2$ the change in sign can occur, both with and without
the $g_3$ contribution and thus is not
just a feature of the NLL contribution at this angle.

The results in fig.\ 4 imply that the relative importance of the NLL
contributions to the
hadronic cross section depend on which regions in $\eta$
are weighted most heavily by the convolution with the parton densities.
With equivalent parton luminosities, if $\eta \approx 0.5$, then we
expect the $gg$ contribution to be comparable to that of the
$q\overline q$ channel while if 
$\eta \approx 2$ then we expect the $q \overline q$
channel to dominate.  Thus the parton luminosity in the $x$ space
probed by the hadronic cross section determines the 
ultimate importance of the NLL contribution.  
A comparison with previous results obtained by integrating 
over the angle $\theta^*$ \cite{LSN1,LSN2}
is inconclusive since here we work at fixed $\theta^*$. Finally we note
that previously \cite{LSN1,LSN2,HERAB,sv2}
different cutoffs were chosen in the $q\overline q$ and
$gg$ channels to match the order-by-order contributions, as described at the
end of the previous section, whereas here we keep
the same $s_{\rm cut}$ for both channels. 

The hadronic cross section calculated to NLL is
\begin{eqnarray}
\sigma^{\rm NLL}(S,m^2) =  \sum_{ij} \int_{\tau_0}^1 d\tau
\int_\tau^1 \frac{dx}{x}
f_i^{h_1}(x,\mu^2) f_j^{h_2}(\frac{\tau}{x},\mu^2)
\sigma_{ij}(\tau S, m^2),
\end{eqnarray}
where $\sigma_{ij}(\tau S, m^2)$ is the partonic cross section, eq.\ (3.2) and
$\tau_0 = (m + \sqrt{m^2 + s_{\rm cut}})^2/S$.  We evaluate the parton 
densities at $\mu^2 = Q^2$.  
Since the parton densities are only available at fixed order, 
the application to a resummed cross section
introduces some uncertainty.  We have used the MRS
D$-^\prime$ DIS densities
for the $q \overline q$ DIS channel and the MRS D$-^\prime$ $\overline{\rm MS}$
densities \cite{mrs2,PDFLIB2} for both channels 
in this scheme\footnote{The $x$ dependence of the
parton densities in the two schemes is very similar.}.  Note that $\Lambda_5 
= 0.1559$ GeV for both sets.  The result is not 
strongly dependent
on the parton distributions in the $x$ range probed at this
energy, $x \sim 0.2$ at $y=0$ for $m=175$ GeV/$c^2$,
entering primarily through the value of $\Lambda_5$ since different
sets of parton densities have a different $\Lambda_5$, changing $s_{4, 
{\rm min}}$ as well as the value of $\alpha_s$.  Taking
$\alpha_s$ to only one loop, as in \cite{LSN1}, increases $\alpha_s$ 
and thus the hadronic cross section.  For a more complete discussion
of the parton density and $\alpha_s$ dependence, see \cite{sv2}. 

In fig.\ 5(a) we plot the $t \overline t$ 
production cross section at $\theta^* = 90^\circ$ as a function of 
top quark mass at the Fermilab Tevatron with $\sqrt{S} = 1.8$ TeV
in the DIS and $\overline{\rm MS}$ 
schemes for the $q \overline
q$ channel and the $\overline{\rm MS}$ scheme for the $gg$ channel
with $s_{\rm cut} = 10s_{4, {\rm min}}$ and $Q^2 = m^2$.
The total $\overline{\rm MS}$ cross section at $\theta^* = 90^\circ$ 
is 5.3 pb at $m=175$ GeV/$c^2$, similar to the angle-integrated
results of Ref.\ \cite{bc}.  Previous NLO results
 showed that for $m=100$ GeV/$c^2$, even though
$\eta_{\rm max} = 80$ the total hadronic cross section was obtained 
already for $\eta \approx 3$ ({\it c.f.}\ fig.1 in \cite{LSN2}).  
When $m=175$ GeV/$c^2$, $\eta_{\rm max} = 26$.
Comparing figs.\ 4a and 5a, we can infer that in this case the important 
$\eta$ region is $\eta \approx 0.3$. At this $\eta$
the $q\overline q$ luminosity is six times larger than the $gg$ luminosity.

To quantify the enhancement in the $t \overline t$ cross section at fixed
angle due to the NLL
terms, in fig.\ 5(b) we show our results with the same values of
$s_{\rm cut}$ and $Q^2$ but without the NLL contributions, {\it i.e.}\ no $g_3$
component.  The NLL contribution increases with top mass at this
energy.  The DIS cross section is enhanced between 25 and 30\% as the top
mass increases from 140 to 200 GeV/$c^2$.  The enhancement is somewhat larger
for the $\overline{\rm MS}$ scheme, between 36 and 43\% in the $q \overline q$
channel and 47 to 55\% in the $gg$ channel.  The change in the amount of
enhancement is, in part, due to the energy dependence of the $g_3$
contribution.  For our choice of $s_{\rm cut}$,
the $t \overline t$ cross section at $m=175$ GeV/$c^2$ without the NLL $g_3$
contribution is 3.7 pb.  
Thus the NLL terms enhance the total $\overline{\rm MS}$
cross section at $\theta^* = 90^\circ$
by 43\% at this mass and value of $s_{\rm cut}$ in our calculation. 
The relatively large contribution from the NLL terms verifies that the lower
$\eta$ region is most important.

We have further investigated the $s_{\rm cut}$ 
and scale dependence of our calculation.
In Tables 1-3 we give the numerical values of the hadronic top
production cross sections at LL and NLL with $s_{\rm cut} = 10s_{4, {\rm
min}}$, $5s_{4, {\rm min}}$ and 20$s_{4, {\rm min}}$ respectively.  
For each value of
$s_{\rm cut}$ we show the results for $Q^2 = m^2$, $4m^2$ and $m^2/4$.
In the tables we have also indicated the equivalent value of the cutoff
$\mu_0$ defined in \cite{LSN1} in both the DIS and $\overline{\rm MS}$ schemes.
Note that in Tables 1 and 3, the scale
dependence of the $gg$ channel is weak compared to the $q \overline q$
contributions.  When $s_{\rm cut} = 10 s_{4, {\rm min}}$, the $gg$ cross
section changes 41\% between $Q^2 = m^2/4$ and $4m^2$ at $m=175$ GeV/$c^2$.
In contrast, the $q \overline q$ $\overline{\rm MS}$ cross section
changes 47\% and the DIS cross section by 65\%.  At the same mass but with
$s_{\rm cut} = 20 s_{4, {\rm min}}$,  the $gg$ cross section changes 10\%
between the highest and lowest scales while the $q \overline q$ $\overline{\rm
MS}$ cross section changes 35\% and the DIS cross section 52\%.  On the other
hand, for the lowest $s_{\rm cut}$ the $gg$ cross section has the largest scale
dependence, more than a factor of two change while the $\overline{\rm MS}$
cross section changes 53\% in the $q \overline q$ channel.

Note that for a fixed value of $s_{\rm cut}$, the $gg$ cross section is
actually largest when $Q^2 = 4m^2$ than at lower $Q^2$.
This result is counterintuitive
since one generally expects that increasing the scale decreases the cross
section.  This is true in the $q \overline q$ channel and also  
in the $gg$ channel when $\eta > 1$.  At fixed $s_4$, the exponents are larger
for lower values of the scale due to the running of $\alpha_s$.  
However, the difference in the
factors $f_{gg,1}$ and $f_{gg,2}$ at low $\eta$ shifts the relative weights
of $d\overline{F}_{gg,I}^{(0)}/ds_4$ and  $d\overline{F}_{gg,II}^{(0)}/ds_4$ 
so that the partonic cross section is largest for the largest scale studied,
$Q^2 = 4m^2$, when $\eta<0.5$.  In addition, changing the scale for fixed
$s_{\rm cut}/s_{4, {\rm min}}$ effectively changes $s_{\rm cut}$ in eq.\ (3.2)
since $s_{4, {\rm min}} = 2m^2 \Lambda/Q$.  Therefore increasing $Q^2$
correspondingly increases the effective $s_4$ integration region.  
This latter effect seems
to be the most important because the largest scale also produces the largest
$gg$ contribution to the cross section when only the LL terms are considered.

When $s_{\rm cut}= 5s_{4, {\rm min}}$, the gluon contribution is enhanced and
actually dominates for $Q^2 = m^2$ and $4m^2$, seen in Table 2. 
This is not surprising
since at $s_{\rm cut} = 5 s_{4, {\rm min}}$, 
$\mu_0 \approx 0.1m$ in the DIS scheme and $0.2m$ in the $\overline{
\rm MS}$ scheme when $Q^2 = m^2$.  
At such low values of $\mu_0$ in the $gg$ channel, the $gg$
cross section can become large, see {\it e.g.}\ figs.\ 12-14 in \cite{LSN1}.
Additionally, as $s_{\rm cut}$ is reduced, all the contributions to the
hadronic cross section increase because as $s_{\rm cut}$ approaches $s_{4, {\rm
min}}$ the exponents grow more rapidly.  The fastest growth of the exponents
in this region occurs when $Q^2 = m^2/4$.

A comparison of the NLL and LL results in the tables helps to clarify where the
NLL enhancements are largest.  As discussed above, the enhancement for a fixed
$s_{\rm cut}$ is largest for $Q^2 = 4m^2$ since the larger scale increases the
effective $s_4$ probed at larger $Q^2$.  The enhancement is also
larger for lower
$s_{\rm cut}$ due to the running of the coupling constant.  Finally we note
that the NLL enhancement increases with mass since $t \overline t$ production
is closer to threshold at fixed energy with larger masses.

We conclude our discussion of top production
by noting that the scheme dependence is rather 
strong, especially for lower $s_{\rm cut}$.
The scheme dependence arises
because $f_{q \overline q}^{\overline {\rm MS}} > f_{q \overline q}^{\rm DIS}$,
particularly near $s_{4, {\rm min}}$, see figs.\ 1 and 2 and also {\it 
e.g.}\ figs.\ 12 and 13 in Ref.\ \cite{LSN1}. 
For $s_{\rm cut}=10s_{4,{\rm min}}$ and $Q^2=m^2$, 
the $\overline{\rm MS}$ cross
section is 62-71\% larger than the DIS cross section at NLL and 49-55\% larger
at LL.  The scheme dependence is largest for
small $s_{\rm cut}$ and large $Q^2$ because smaller values of $s_{\rm cut}$
are increasingly sensitive to the upper limit of the $\xi$ integral in 
$E_{ab}(g_1)$, see the appendix.   The scheme
dependence increases slightly with mass at fixed energy. 
Thus close to threshold the scheme dependence is unavoidable. 
Even though the partonic cross section is only weakly scheme dependent for
$\eta > 1$, as seen in fig.\ 4, some scheme dependence will remain 
until new sets of parton densities 
which have incorporated resummation effects before 
being fitted to data are available. In the absence of such densities we
favor the $\overline{\rm MS}$ results as they have a more reliable 
theoretical basis and the $q \overline q$ and $gg$ channels can be treated
consistently only in this scheme.

Since $\alpha_s$ depends strongly on the heavy quark mass scale,
we repeat our calculations for bottom quark production
with $m = 4.75$ GeV/$c^2$ at $\sqrt{s} = 9.51$ GeV.  The smaller quark mass and
the larger $\Lambda_4$, $0.23$ GeV for the MRS D$-'$ distributions,
result in a reduced range in $s_4/2m^2$ before the
exponents diverge, as shown in figs.\ 6-8.  The exponents tend to be somewhat
smaller for the bottom quark except near $s_{4, {\rm min}}$ due to the faster
running of the coupling constant. The running of $\alpha_s$ also produces
faster growth of the exponents at lower $Q^2$.  Thus the scale dependence is
also stronger for $b \overline b$ production. 
We note that the energy
dependence of $g_3$ is stronger for the lighter quark mass.  While
$E_{gg}(\lambda_1)$ is smaller near threshold, it increases faster with energy
than at the top quark mass. 
In general, the bottom quark cross
section is more sensitive to $s_{\rm cut}$ and $Q^2$ than the much more massive
top quarks, see also \cite{HERAB,sv2}.  

The partonic cross sections with
$s_{\rm cut} = 1.4s_{4, {\rm min}}$ and $Q^2 = m^2$ are shown with and without
the NLL $g_3$ terms in fig.\ 9. This value of $s_{\rm cut}$ corresponds to
$\mu_0 \approx 0.37m$ and
$0.51m$ in the DIS and  $\overline {\rm MS}$ schemes, somewhat larger than 
those in \cite{HERAB,sv2}.
Varying $s_{\rm cut}$ we find that for $\eta < 1$,
the partonic cross section can change by an order of magnitude near $\eta \sim
0.1$ and a factor of two to three at $\eta \sim 0.5$.  The largest variation
is in the $gg$ channel. Of course the parton luminosity in the
$x$ range probed at a given energy
determines the sensitivity of the hadronic cross section.  At higher
energies, away from threshold, the sensitivity to $s_{\rm cut}$ decreases.
Similar trends can be seen in charm 
production although for $c \overline c$ production, $s_{4, {\rm min}}$ 
already corresponds to the $\mu_0$ values used in \cite{sv2}, suggesting that
a stable cross section is even more difficult to obtain for charm production.

In fig.\ 10 we show the $b \overline b$ production cross section at $\theta^* =
90^\circ$ as a 
function of beam momentum for $pp$ interactions with $s_{\rm cut} \approx
1.4s_{4, {\rm min}}$ and $Q^2 = m^2$, with and without the NLL contributions.
As expected, $gg$ fusion is dominant \cite{HERAB,sv2}.  These
energies, $20 \leq \sqrt{S} \leq 45$ GeV, correspond 
to $3.4 \leq \eta_{\rm max} \leq 21.4$.  
At the HERA-B energy, $\sqrt{S} = 39.2$ GeV, we find a total
$\overline{\rm MS}$ cross section at $\theta^* = 90^\circ$
of 8.6 nb at LL and 19.1 nb at NLL, more than a 
factor of two enhancement.  Note that the $b \overline b$ cross
section is somewhat smaller than found previously \cite{HERAB,sv2}, 
in part because we are only considering a fixed angle.
  
The sensitivity to $s_{\rm cut}$ and the scale is detailed in Tables 4-6 for
$s_{\rm cut}= 1.2$, 1.4, and 
1.6$s_{4, {\rm min}}$ and $Q^2 = m^2$, $4m^2$ and $m^2/4$.  The same
trends can be observed as we remarked upon for $t \overline t$ production
although the enhancements are generally larger for $b \overline b$ production.
We note that for the values of $s_{\rm cut}$ chosen here, the NLL
enhancement and scheme dependence decrease with increasing energy since the
higher energy probes the region where the corrections are less important.
There seems to be
a general decrease of the scheme dependence with energy for the bottom quarks.
However, it is difficult to fairly
compare the scheme dependence of top and bottom production since our values of
$s_{\rm cut}$
do not correspond to the same $\mu_0/m$ ratio in both cases.  Our previous work
on LL resummation favored a fixed ratio to describe production of all heavy
quarks \cite{HERAB,sv2}.
Thus, while our bottom quark results are illustrative only, they clearly
indicate that a full NLL calculation including all angles is needed to clarify
bottom quark resummation.

\mysection{Conclusions}

We have investigated the numerical importance of the NLL terms in the
resummation of subleading soft gluon contributions near threshold for 
heavy quark production. We have shown that, for the resummation method of 
\cite{LSN1} and at $\theta^* = 90^\circ$ these contributions
are either numerically small or there are cancellations between
them.  Therefore the inclusion of the NLL terms leads to only a moderate change
in the top quark production cross section at $\theta^* = 90^\circ$.
Bottom production is much more sensitive
to $s_{\rm cut}$, making definitive statements about the resummed cross
sections more difficult, particularly in a model without a calculable
perturbative cutoff.\\[2ex]

We would like to thank G. Sterman and E. Laenen for helpful discussions.  
R. Vogt and N.  Kidonakis would like to thank Brookhaven National 
Laboratory for hospitality.
R. Vogt also thanks SUNY at Stony Brook for hospitality.  
J. Smith would like to
thank the Alexander von Humboldt Stiftung for an award allowing him to spend
his sabbatical year at DESY.

\setcounter{section}{0}
\renewcommand{\thesection}{\Alph{section}}
\mysection{Appendix}

Here we present the analytical results for the $g_1$, $g_2$ and $g_3$ (where
appropriate) integrals.
As previously discussed, we use the two-loop running coupling constant,
\begin{eqnarray}
 && \alpha_s(\xi) = \frac{1}{a \ln \xi}
+ \frac{b}{a} \frac{\ln(\ln \xi)}{\ln^2(\xi)} \,, \\
 && a = \frac{11C_A - 4T_f n_{\rm f}}{12\pi} \,,  \nonumber \\
&& b = -6 \frac{17 C_A^2 - (6C_F + 10C_A)T_fn_{\rm f}}
{(11 C_A - 4 T_f n_{\rm f})^2}\,, \nonumber
\end{eqnarray}
with $C_A = 3$, $C_F = 4/3$ and $T_f = 1/2$. It is obvious that when 
$\xi < 1$, $\ln\xi$ 
is negative so that $\ln(\ln\xi)$ has a cut and needs a precise definition. We 
use the cutoff on the $\omega'$ variable to stop the integration at
that point.  

In general,
\begin{eqnarray}
\lefteqn{E_{ab}^{\rm sch} = \int_{\omega_0}^1\frac{d\omega'}{\omega'}
\Big\{\int_{\xi_L}^{\xi_U^{\rm sch}} \frac{d\xi}{\xi}\,
 \Big[ \gamma_{ab} \Big( \alpha_s(\xi) 
+ \frac{K}{2\pi} \alpha^2_s(\xi) \Big) \Big] 
 - \delta^{\rm sch}_{ab} \alpha_s
\Big( \frac{\omega' Q^2}{\Lambda^2}\Big)} \nonumber \\
& & \!\!\!\!\!\!\!\! -\Big\{
 \lambda_{i, ab} 
\Big[\alpha_s \Big(\frac{\omega'^2 Q^2}{\Lambda^2}\Big),\theta^*=90^\circ
\Big]  
 + \lambda_{i, ab}^* 
\Big[\alpha_s \Big(\frac{\omega'^2 Q^2}{\Lambda^2}\Big),\theta^*=90^\circ
\Big]  
\, \Big\} \Big\} \, \, 
\end{eqnarray}
where $\xi_L = \omega'^2 Q^2/\Lambda^2$, $\xi_U^{\rm DIS} =  \omega' 
Q^2/\Lambda^2$, $\xi_U^{\overline{\rm MS}} = Q^2/\Lambda^2$, $\gamma_{q
\overline q} = 2C_F/\pi$, $\gamma_{gg} = 2C_A/\pi$ and $\delta^{\rm sch}_{ab} =
3C_F/2\pi$ for $q \overline q$ in the DIS scheme and 0 otherwise.
We will give the results for single integrals over $\omega'$ first.
For convenience, we define 
\begin{eqnarray} 
L_0 = \ln \Big( \frac{Q^2}{\Lambda^2} \Big)\, \, ; \,\,\,\,\,\,
L_1 = \ln \Big( \frac{\omega_0 Q^2}{\Lambda^2} \Big)\, \, ; \,\,\,\,\,\,
L_2 = \ln \Big( \frac{\omega_0^2 Q^2}{\Lambda^2} \Big)\, \, .
\end{eqnarray}
Note that the contributions dependent on $L_2$ set the lower limit on $s_{\rm 
cut}$ since the value of $\omega_0$ where $\omega_0 \sim \Lambda/Q$ is the
lowest $\omega_0$ at which the integrals diverge.

Then, in the DIS scheme, the $g_2$ result is
\begin{eqnarray} E_{q \overline q}^{\rm DIS}(g_2) = 
- \frac{3}{2} \frac{C_F}{\pi a} \Big[ \ln L_0 - \frac{b}{L_0}(\ln L_0 + 1)
- \ln L_1 + \frac{b}{L_1}(\ln L_1 + 1) \Big] \, \, .
\end{eqnarray}
For the terms in $\lambda_{i, ab}$ which only contain an $s_4'$ dependence in
the running coupling constant, the $g_3$ integral can also
be done analytically.  In these cases,
\begin{eqnarray} E_{ab}(g_3) \propto \frac{1}{2 a} \Big[ \ln L_0 - 
\frac{b}{L_0}(\ln L_0 + 1)
- \ln L_2 + \frac{b}{L_2}(\ln L_2 + 1) \Big] \, \, .
\end{eqnarray}
Thus the $g_3$ contributions diverge at a larger value of $\omega_0$ than do
the $g_2$.

After the $\xi$ integral, the $g_1$ exponent is
\begin{eqnarray}
E_{ab}^{\rm sch}(g_1) & = & \frac{\gamma_{ab}}{a} \int_{\omega_0}^1 \frac{d
\omega'}{\omega'} \Big[ \Big\{ \ln(\ln \xi) - \frac{b}{\ln \xi} (\ln(\ln \xi)
+ 1) \Big\} - \frac{K}{2 \pi a \ln \xi} \Big\{ 1 \\
& & \!\!\!\!\!\!\!\! 
+ \frac{b}{2 \ln \xi}(2 \ln(\ln \xi) + 1) + \frac{b^2}{27 \ln^2 \xi}
(9 \ln^2 (\ln \xi) + 6 \ln (\ln \xi) + 2) \Big\} \Big] \Big|_{\xi_L}^{\xi_U^{\rm
sch}} \, \, . \nonumber
\end{eqnarray}
In the DIS scheme the final result is
\begin{eqnarray}
\lefteqn{E_{ab}^{\rm DIS}(g_1) = \frac{\gamma_{ab}}{a} \Big[ \frac{1}{2} 
L_0 ( \ln L_0 - 1) - \frac{b}{4} \ln L_0 (\ln L_0 + 2)}  \nonumber \\
& & - L_1 ( \ln L_1 - 1) + \frac{b}{2} \ln L_1 (\ln L_1 + 2) \nonumber \\
& & + \frac{1}{2} L_2 ( \ln L_2 - 1) - \frac{b}{4} \ln L_2 (\ln L_2 + 2) \\
& & -\frac{K}{2 \pi a} \Big\{ \frac{1}{2} \ln L_0 - \frac{b}{4 L_0} (2 \ln L_0
+3) - \frac{b^2}{216 L_0^2} (18 \ln^2 L_0 + 30 \ln L_0 + 19) \nonumber \\
& & - \ln L_1 + \frac{b}{2 L_1} (2 \ln L_1
+3) + \frac{b^2}{108 L_1^2} (18 \ln^2 L_1 + 30 \ln L_1 + 19) \nonumber \\
& & +  \frac{1}{2} \ln L_2 - \frac{b}{4 L_2} (2 \ln L_2
+3) - \frac{b^2}{216 L_2^2} (18 \ln^2 L_2 + 30 \ln L_2 + 19) \Big\} \Big] \, \,
. \nonumber
\end{eqnarray}
The result is somewhat simpler in the $\overline{\rm MS}$ scheme since
$\xi_U^{\overline{\rm MS}}$ does not depend on $\omega'$.  In this case,
\begin{eqnarray}
\lefteqn{E_{ab}^{\overline{\rm MS}}(g_1) = 
\frac{\gamma_{ab}}{a} \Big[ -\ln \omega_0 \Big\{ \ln L_0 - \frac{b}{L_0} (\ln
L_0 + 1) \Big\} } \nonumber \\
& & - \frac{1}{2} L_0 ( \ln L_0 - 1) + \frac{b}{4} \ln L_0 (\ln L_0 + 2)  
\nonumber \\
& & + \frac{1}{2} L_2 ( \ln L_2 - 1) - \frac{b}{4} \ln L_2 (\ln L_2 + 2) \\
& & -\frac{K}{2 \pi a} \Big\{ -\frac{\ln \omega_0}{L_0}
\Big[ 1 + \frac{b}{2L_0}(2 \ln L_0 + 1) + \frac{b^2}{27L_0^2} (9 \ln^2 L_0
+ 6 \ln L_0 + 2) \Big] \nonumber \\
& & -\frac{1}{2} \ln L_0 + \frac{b}{4 L_0} (2 \ln L_0
+3) + \frac{b^2}{216 L_0^2} (18 \ln^2 L_0 + 30 \ln L_0 + 19) \nonumber \\
& & +  \frac{1}{2} \ln L_2 - \frac{b}{4 L_2} (2 \ln L_2
+3) - \frac{b^2}{216 L_2^2} (18 \ln^2 L_2 + 30 \ln L_2 + 19) \Big\} \Big] \, \,
. \nonumber
\end{eqnarray}

\newpage

\begin{table}[htb]
\begin{tabular}{|c|c|c|c|c|c|c|} \hline
 & \multicolumn{3}{c|}{NLL} & \multicolumn{3}{c|}{LL} \\ \hline
$m$ (GeV/$c^2$) & $\sigma_{q \overline q}^{\rm DIS}$ (pb) &  $\sigma_{q 
\overline q}^{\overline{\rm MS}}$ (pb) & $\sigma_{gg}$ (pb) & $\sigma_{q 
\overline q}^{\rm DIS}$ (pb) &  $\sigma_{q 
\overline q}^{\overline{\rm MS}}$ (pb) & $\sigma_{gg}$ (pb) \\ \hline
 & \multicolumn{6}{c|}{$Q^2 = m^2 \Rightarrow \mu_0^{\rm DIS} = 0.13m,
\mu_0^{\overline{\rm MS}} = 0.26m$} \\ \hline
 150 & 4.91 &  8.04 & 3.99 & 3.91 & 5.85 & 2.68 \\ \hline
 155 & 4.21 &  6.94 & 3.21 & 3.32 & 5.01 & 2.15 \\ \hline
 160 & 3.63 &  5.99 & 2.61 & 2.87 & 4.33 & 1.75 \\ \hline
 165 & 3.13 &  5.20 & 2.12 & 2.47 & 3.74 & 1.40 \\ \hline
 170 & 2.71 &  4.51 & 1.73 & 2.12 & 3.23 & 1.14 \\ \hline
 175 & 2.34 &  3.92 & 1.41 & 1.84 & 2.80 & 0.93 \\ \hline
 180 & 2.03 &  3.41 & 1.15 & 1.59 & 2.43 & 0.75 \\ \hline
 185 & 1.77 &  2.98 & 0.94 & 1.37 & 2.11 & 0.62 \\ \hline
 190 & 1.53 &  2.59 & 0.77 & 1.19 & 1.84 & 0.50 \\ \hline
 195 & 1.33 &  2.27 & 0.63 & 1.03 & 1.60 & 0.41 \\ \hline
 200 & 1.16 &  1.99 & 0.52 & 0.90 & 1.39 & 0.34 \\ \hline
 & \multicolumn{6}{c|}{$Q^2 = 4m^2 \Rightarrow \mu_0^{\rm DIS} = 0.19m,
\mu_0^{\overline{\rm MS}} = 0.41m$} \\ \hline
 150 & 3.83 &  6.60 & 4.94 & 3.06 & 4.74 & 3.11 \\ \hline
 155 & 3.30 &  5.71 & 3.95 & 2.61 & 4.07 & 2.49 \\ \hline
 160 & 2.83 &  4.92 & 3.22 & 2.24 & 3.50 & 2.01 \\ \hline
 165 & 2.44 &  4.27 & 2.61 & 1.93 & 3.03 & 1.62 \\ \hline
 170 & 2.10 &  3.69 & 2.13 & 1.65 & 2.61 & 1.32 \\ \hline
 175 & 1.82 &  3.21 & 1.74 & 1.43 & 2.26 & 1.08 \\ \hline
 180 & 1.56 &  2.78 & 1.42 & 1.23 & 1.95 & 0.88 \\ \hline
 185 & 1.36 &  2.42 & 1.16 & 1.06 & 1.69 & 0.71 \\ \hline
 190 & 1.18 &  2.11 & 0.96 & 0.92 & 1.47 & 0.59 \\ \hline
 195 & 1.03 &  1.84 & 0.79 & 0.80 & 1.28 & 0.48 \\ \hline
 200 & 0.89 &  1.61 & 0.65 & 0.69 & 1.11 & 0.39 \\ \hline
 & \multicolumn{6}{c|}{$Q^2 = m^2/4 \Rightarrow \mu_0^{\rm DIS} = 0.094m,
\mu_0^{\overline{\rm MS}} = 0.16m$} \\ \hline
 150 & 6.20 &  9.53 & 3.51 & 4.97 & 7.13 & 2.51 \\ \hline
 155 & 5.36 &  8.28 & 2.84 & 4.26 & 6.12 & 2.03 \\ \hline
 160 & 4.62 &  7.17 & 2.29 & 3.67 & 5.29 & 1.62 \\ \hline
 165 & 4.00 &  6.24 & 1.85 & 3.16 & 4.57 & 1.31 \\ \hline
 170 & 3.46 &  5.41 & 1.51 & 2.73 & 3.97 & 1.06 \\ \hline
 175 & 3.00 &  4.72 & 1.23 & 2.36 & 3.44 & 0.86 \\ \hline
 180 & 2.61 &  4.11 & 1.00 & 2.04 & 2.98 & 0.70 \\ \hline
 185 & 2.27 &  3.59 & 0.82 & 1.77 & 2.60 & 0.57 \\ \hline
 190 & 1.99 &  3.16 & 0.68 & 1.54 & 2.26 & 0.47 \\ \hline
 195 & 1.73 &  2.75 & 0.55 & 1.34 & 1.97 & 0.38 \\ \hline
 200 & 1.51 &  2.42 & 0.46 & 1.16 & 1.72 & 0.31 \\ \hline
\end{tabular}
\caption[]{The hadronic $t \overline t$ production cross sections 
in $p \overline p$ collisions at $\sqrt{S} = 1.8$ TeV for
$s_{\rm cut} = 10s_{4, {\rm min}}$ and $\theta^* = 90^\circ$.  
The effective $\mu_0$ from \cite{LSN1}
is given for each $Q^2$.}
\end{table}

\newpage
\begin{table}[htb]
\begin{tabular}{|c|c|c|c|c|c|c|} \hline
 & \multicolumn{3}{c|}{NLL} & \multicolumn{3}{c|}{LL} \\ \hline
$m$ (GeV/$c^2$) & $\sigma_{q \overline q}^{\rm DIS}$ (pb) &  $\sigma_{q 
\overline q}^{\overline{\rm MS}}$ (pb) & $\sigma_{gg}$ (pb) & $\sigma_{q 
\overline q}^{\rm DIS}$ (pb) &  $\sigma_{q 
\overline q}^{\overline{\rm MS}}$ (pb) & $\sigma_{gg}$ (pb) \\ \hline
 & \multicolumn{6}{c|}{$Q^2 = m^2 \Rightarrow \mu_0^{\rm DIS} = 0.094m,
\mu_0^{\overline{\rm MS}} = 0.21m$} \\ \hline
 150 & 5.43 & 10.81 & 17.31 & 4.13 & 7.07 &  9.67 \\ \hline
 155 & 4.65 &  9.29 & 14.13 & 3.55 & 6.09 &  7.89 \\ \hline
 160 & 4.01 &  8.08 & 11.66 & 3.05 & 5.25 &  6.44 \\ \hline
 165 & 3.47 &  7.02 &  9.54 & 2.63 & 4.55 &  5.27 \\ \hline
 170 & 2.99 &  6.10 &  7.93 & 2.25 & 3.93 &  4.30 \\ \hline
 175 & 2.59 &  5.32 &  6.53 & 1.95 & 3.41 &  3.54 \\ \hline
 180 & 2.25 &  4.63 &  5.38 & 1.69 & 2.96 &  2.90 \\ \hline
 185 & 1.95 &  4.04 &  4.47 & 1.46 & 2.58 &  2.41 \\ \hline
 190 & 1.70 &  3.56 &  3.68 & 1.26 & 2.24 &  1.99 \\ \hline
 195 & 1.48 &  3.09 &  3.06 & 1.10 & 1.96 &  1.65 \\ \hline
 200 & 1.28 &  2.71 &  2.55 & 0.96 & 1.71 &  1.37 \\ \hline
 & \multicolumn{6}{c|}{$Q^2 = 4m^2 \Rightarrow \mu_0^{\rm DIS} = 0.13m,
\mu_0^{\overline{\rm MS}} = 0.33m$} \\ \hline
 150 & 4.15 &  8.80 & 26.50 & 3.20 & 5.68 & 14.01 \\ \hline
 155 & 3.55 &  7.58 & 21.69 & 2.73 & 4.86 & 11.32 \\ \hline
 160 & 3.06 &  6.55 & 17.85 & 2.34 & 4.19 &  9.22 \\ \hline
 165 & 2.63 &  5.68 & 14.66 & 2.01 & 3.63 &  7.57 \\ \hline
 170 & 2.28 &  4.95 & 12.03 & 1.73 & 3.12 &  6.26 \\ \hline
 175 & 1.96 &  4.29 & 10.03 & 1.49 & 2.71 &  5.13 \\ \hline
 180 & 1.70 &  3.74 &  8.24 & 1.28 & 2.34 &  4.23 \\ \hline
 185 & 1.47 &  3.26 &  6.86 & 1.11 & 2.04 &  3.51 \\ \hline
 190 & 1.28 &  2.85 &  5.65 & 0.96 & 1.77 &  2.88 \\ \hline
 195 & 1.11 &  2.48 &  4.74 & 0.83 & 1.54 &  2.38 \\ \hline
 200 & 0.96 &  2.17 &  3.89 & 0.72 & 1.34 &  1.98 \\ \hline
 & \multicolumn{6}{c|}{$Q^2 = m^2/4 \Rightarrow \mu_0^{\rm DIS} = 0.067m,
\mu_0^{\overline{\rm MS}} = 0.13m$} \\ \hline
 150 & 7.10 & 13.16 & 12.44 & 5.41 &  8.84 &  7.42 \\ \hline
 155 & 6.13 & 11.44 & 10.21 & 4.65 &  7.63 &  6.04 \\ \hline
 160 & 5.30 &  9.91 &  8.35 & 3.99 &  6.58 &  4.91 \\ \hline
 165 & 4.58 &  8.65 &  6.86 & 3.44 &  5.69 &  4.03 \\ \hline
 170 & 3.95 &  7.49 &  5.64 & 2.97 &  4.94 &  3.28 \\ \hline
 175 & 3.44 &  6.57 &  4.65 & 2.57 &  4.29 &  2.71 \\ \hline
 180 & 2.98 &  5.72 &  3.84 & 2.23 &  3.75 &  2.25 \\ \hline
 185 & 2.60 &  5.02 &  3.20 & 1.93 &  3.25 &  1.84 \\ \hline
 190 & 2.27 &  4.40 &  2.65 & 1.68 &  2.84 &  1.52 \\ \hline
 195 & 1.98 &  3.85 &  2.19 & 1.46 &  2.48 &  1.25 \\ \hline
 200 & 1.73 &  3.38 &  1.82 & 1.28 &  2.17 &  1.04 \\ \hline
\end{tabular}
\caption[]{The hadronic $t \overline t$ production cross sections 
in $p \overline p$ collisions at $\sqrt{S} = 1.8$ TeV for
$s_{\rm cut} = 5s_{4, {\rm min}}$ and  $\theta^* = 90^\circ$.  
The effective $\mu_0$ from \cite{LSN1}
is given for each $Q^2$.}
\end{table}

\newpage
\begin{table}[htb]
\begin{tabular}{|c|c|c|c|c|c|c|} \hline
 & \multicolumn{3}{c|}{NLL} & \multicolumn{3}{c|}{LL} \\ \hline
$m$ (GeV/$c^2$) & $\sigma_{q \overline q}^{\rm DIS}$ (pb) &  $\sigma_{q 
\overline q}^{\overline{\rm MS}}$ (pb) & $\sigma_{gg}$ (pb) & $\sigma_{q 
\overline q}^{\rm DIS}$ (pb) &  $\sigma_{q 
\overline q}^{\overline{\rm MS}}$ (pb) & $\sigma_{gg}$ (pb) \\ \hline
 & \multicolumn{6}{c|}{$Q^2 = m^2 \Rightarrow \mu_0^{\rm DIS} = 0.19m,
\mu_0^{\overline{\rm MS}} = 0.33m$} \\ \hline
 150 & 4.40 & 6.30 & 1.61 & 3.61 & 4.90 & 1.24 \\ \hline
 155 & 3.78 & 5.43 & 1.29 & 3.10 & 4.23 & 0.98 \\ \hline
 160 & 3.25 & 4.69 & 1.03 & 2.66 & 3.64 & 0.78 \\ \hline
 165 & 2.80 & 4.06 & 0.83 & 2.28 & 3.13 & 0.63 \\ \hline
 170 & 2.42 & 3.53 & 0.67 & 1.97 & 2.71 & 0.50 \\ \hline
 175 & 2.10 & 3.06 & 0.54 & 1.70 & 2.34 & 0.40 \\ \hline
 180 & 1.82 & 2.67 & 0.44 & 1.47 & 2.03 & 0.33 \\ \hline
 185 & 1.58 & 2.32 & 0.35 & 1.27 & 1.76 & 0.26 \\ \hline
 190 & 1.37 & 2.02 & 0.29 & 1.10 & 1.53 & 0.21 \\ \hline
 195 & 1.19 & 1.76 & 0.23 & 0.96 & 1.33 & 0.17 \\ \hline
 200 & 1.04 & 1.55 & 0.19 & 0.83 & 1.16 & 0.14 \\ \hline
 & \multicolumn{6}{c|}{$Q^2 = 4m^2 \Rightarrow \mu_0^{\rm DIS} = 0.27m,
\mu_0^{\overline{\rm MS}} = 0.52m$} \\ \hline
 150 & 3.55 & 5.36 & 1.70 & 2.91 & 4.11 & 1.24 \\ \hline
 155 & 3.05 & 4.61 & 1.36 & 2.48 & 3.52 & 0.99 \\ \hline
 160 & 2.62 & 3.98 & 1.09 & 2.13 & 3.03 & 0.79 \\ \hline
 165 & 2.25 & 3.44 & 0.87 & 1.83 & 2.61 & 0.63 \\ \hline
 170 & 1.94 & 2.97 & 0.71 & 1.57 & 2.25 & 0.51 \\ \hline
 175 & 1.68 & 2.58 & 0.57 & 1.35 & 1.94 & 0.41 \\ \hline
 180 & 1.45 & 2.24 & 0.46 & 1.17 & 1.68 & 0.33 \\ \hline
 185 & 1.26 & 1.95 & 0.38 & 1.01 & 1.46 & 0.27 \\ \hline
 190 & 1.09 & 1.69 & 0.31 & 0.87 & 1.26 & 0.22 \\ \hline
 195 & 0.95 & 1.48 & 0.25 & 0.75 & 1.09 & 0.18 \\ \hline
 200 & 0.82 & 1.28 & 0.20 & 0.69 & 0.96 & 0.14 \\ \hline
 & \multicolumn{6}{c|}{$Q^2 = m^2/4 \Rightarrow \mu_0^{\rm DIS} = 0.13m,
\mu_0^{\overline{\rm MS}} = 0.21m$} \\ \hline
 150 & 5.27 & 7.10 & 1.57 & 4.40 & 5.68 & 1.25 \\ \hline
 155 & 4.54 & 6.14 & 1.25 & 3.77 & 4.89 & 1.00 \\ \hline
 160 & 3.92 & 5.31 & 0.99 & 3.25 & 4.23 & 0.80 \\ \hline
 165 & 3.39 & 4.62 & 0.80 & 2.80 & 3.65 & 0.64 \\ \hline
 170 & 2.94 & 4.02 & 0.64 & 2.41 & 3.16 & 0.51 \\ \hline
 175 & 2.55 & 3.49 & 0.52 & 2.08 & 2.74 & 0.41 \\ \hline
 180 & 2.22 & 3.05 & 0.42 & 1.81 & 2.38 & 0.33 \\ \hline
 185 & 1.93 & 2.67 & 0.34 & 1.57 & 2.07 & 0.27 \\ \hline
 190 & 1.68 & 2.33 & 0.28 & 1.37 & 1.80 & 0.22 \\ \hline
 195 & 1.47 & 2.04 & 0.23 & 1.18 & 1.57 & 0.18 \\ \hline
 200 & 1.28 & 1.78 & 0.18 & 1.03 & 1.37 & 0.14 \\ \hline
\end{tabular}
\caption[]{The hadronic $t \overline t$ production cross sections 
in $p \overline p$ collisions at $\sqrt{S} = 1.8$ TeV for
$s_{\rm cut} = 20s_{4, {\rm min}}$ and $\theta^* = 90^\circ$.  
The effective $\mu_0$ from \cite{LSN1}
is given for each $Q^2$.}
\end{table}

\newpage
\begin{table}[htb]
\begin{tabular}{|c|c|c|c|c|c|c|} \hline
 & \multicolumn{3}{c|}{NLL} & \multicolumn{3}{c|}{LL} \\ \hline
$p_{\rm lab}$ (GeV/$c$) & $\sigma_{q \overline q}^{\rm DIS}$ (nb) & $\sigma_{q 
\overline q}^{\overline{\rm MS}}$ (nb) & $\sigma_{gg}$ (nb) & $\sigma_{q 
\overline q}^{\rm DIS}$ (nb) &  $\sigma_{q 
\overline q}^{\overline{\rm MS}}$ (nb) & $\sigma_{gg}$ (nb) \\ \hline
 & \multicolumn{6}{c|}{$Q^2 = m^2 \Rightarrow \mu_0^{\rm DIS} = 0.37m,
\mu_0^{\overline{\rm MS}} = 0.51m$} \\ \hline
 200 & 0.0049 & 0.0081 & 0.034 & 0.0021 & 0.0031 & 0.014 \\ \hline
 300 & 0.049  & 0.091  & 0.410 & 0.023  & 0.037  & 0.172 \\ \hline
 400 & 0.169  & 0.326  & 1.54  & 0.083  & 0.140  & 0.658 \\ \hline
 500 & 0.373  & 0.734  & 3.58  & 0.190  & 0.325  & 1.56  \\ \hline
 600 & 0.653  & 1.30   & 6.52  & 0.344  & 0.591  & 2.90  \\ \hline
 700 & 1.00   & 2.01   & 10.44 & 0.537  & 0.927  & 4.64  \\ \hline
 800 & 1.40   & 2.84   & 15.00 & 0.764  & 1.32   & 6.76  \\ \hline
 900 & 1.84   & 3.72   & 20.23 & 1.02   & 1.77   & 9.20  \\ \hline
 1000 &2.31   & 4.70   & 26.18 & 1.30   & 2.25   & 11.86 \\ \hline
 & \multicolumn{6}{c|}{$Q^2 = 4m^2 \Rightarrow \mu_0^{\rm DIS} = 0.52m,
\mu_0^{\overline{\rm MS}} = 0.82m$} \\ \hline
 200 & 0.0036 & 0.0079 & 0.041 & 0.0016 & 0.0029 & 0.015 \\ \hline
 300 & 0.036  & 0.082  & 0.482 & 0.017  & 0.032  & 0.183 \\ \hline
 400 & 0.124  & 0.286  & 1.81  & 0.060  & 0.118  & 0.702 \\ \hline
 500 & 0.273  & 0.636  & 4.32  & 0.138  & 0.269  & 1.69  \\ \hline
 600 & 0.479  & 1.12   & 8.01  & 0.248  & 0.484  & 3.20  \\ \hline
 700 & 0.732  & 1.72   & 12.87 & 0.387  & 0.756  & 5.15  \\ \hline
 800 & 1.02   & 2.41   & 18.86 & 0.551  & 1.07   & 7.60  \\ \hline
 900 & 1.35   & 3.17   & 25.91 & 0.735  & 1.43   & 10.42 \\ \hline
 1000 &1.70   & 4.01   & 33.68 & 0.941  & 1.83   & 13.69 \\ \hline
 & \multicolumn{6}{c|}{$Q^2 = m^2/4 \Rightarrow \mu_0^{\rm DIS} = 0.26m,
\mu_0^{\overline{\rm MS}} = 0.32m$} \\ \hline
 200 & 0.0041 & 0.0054 & 0.026 & 0.0019 & 0.0022 & 0.012 \\ \hline
 300 & 0.050  & 0.076  & 0.354 & 0.024  & 0.033  & 0.167 \\ \hline
 400 & 0.188  & 0.300  & 1.37  & 0.096  & 0.139  & 0.665 \\ \hline
 500 & 0.433  & 0.713  & 3.20  & 0.232  & 0.341  & 1.59  \\ \hline
 600 & 0.783  & 1.30   & 5.78  & 0.429  & 0.640  & 2.94  \\ \hline
 700 & 1.21   & 2.04   & 9.11  & 0.682  & 1.03   & 4.69  \\ \hline
 800 & 1.72   & 2.92   & 12.96 & 0.985  & 1.49   & 6.79  \\ \hline
 900 & 2.27   & 3.90   & 17.37 & 1.33   & 2.02   & 9.16  \\ \hline
 1000 &2.87   & 4.95   & 22.29 & 1.70   & 2.60   & 11.81 \\ \hline
\end{tabular}
\caption[]{The hadronic $b \overline b$ production cross sections in
$pp$ interactions for $s_{\rm cut} = 1.4s_{4, {\rm min}}$, $m = 4.75$ 
GeV/$c^2$ and $\theta^* = 90^\circ$.}
\end{table}

\newpage
\begin{table}[htb]
\begin{tabular}{|c|c|c|c|c|c|c|} \hline
 & \multicolumn{3}{c|}{NLL} & \multicolumn{3}{c|}{LL} \\ \hline
$p_{\rm lab}$ (GeV/$c$) & $\sigma_{q \overline q}^{\rm DIS}$ (nb) & $\sigma_{q 
\overline q}^{\overline{\rm MS}}$ (nb) & $\sigma_{gg}$ (nb) & $\sigma_{q 
\overline q}^{\rm DIS}$ (nb) &  $\sigma_{q 
\overline q}^{\overline{\rm MS}}$ (nb) & $\sigma_{gg}$ (nb) \\ \hline
 & \multicolumn{6}{c|}{$Q^2 = m^2 \Rightarrow \mu_0^{\rm DIS} = 0.34m,
\mu_0^{\overline{\rm MS}} = 0.49m$} \\ \hline
 200 & 0.0094 & 0.020 & 0.248 & 0.0029 & 0.0049 & 0.051 \\ \hline
 300 & 0.087  & 0.206 & 2.75  & 0.029  & 0.055  & 0.594 \\ \hline
 400 & 0.287  & 0.700 & 9.82  & 0.103  & 0.199  & 2.19  \\ \hline
 500 & 0.617  & 1.54  & 22.43 & 0.232  & 0.452  & 5.06  \\ \hline
 600 & 1.06   & 2.67  & 40.53 & 0.413  & 0.811  & 9.21  \\ \hline
 700 & 1.59   & 4.05  & 63.18 & 0.639  & 1.26   & 14.67 \\ \hline
 800 & 2.20   & 5.62  & 89.85 & 0.902  & 1.78   & 20.84 \\ \hline
 900 & 2.85   & 7.32  & 120.6 & 1.20   & 2.37   & 28.00 \\ \hline
 1000 &3.56   & 9.18  & 154.2 & 1.51   & 2.99   & 36.60 \\ \hline
 & \multicolumn{6}{c|}{$Q^2 = 4m^2 \Rightarrow \mu_0^{\rm DIS} = 0.48m,
\mu_0^{\overline{\rm MS}} = 0.77m$} \\ \hline
 200 & 0.0063 & 0.017 & 0.301 & 0.0019 & 0.0042 & 0.057 \\ \hline
 300 & 0.059  & 0.171 & 3.41  & 0.020  & 0.045  & 0.665 \\ \hline
 400 & 0.193  & 0.575 & 12.56 & 0.071  & 0.159  & 2.51  \\ \hline
 500 & 0.416  & 1.25  & 29.35 & 0.159  & 0.356  & 5.96  \\ \hline
 600 & 0.716  & 2.16  & 54.00 & 0.284  & 0.635  & 11.10 \\ \hline
 700 & 1.08   & 3.30  & 85.96 & 0.442  & 0.985  & 17.73 \\ \hline
 800 & 1.49   & 4.58  & 125.7 & 0.626  & 1.39   & 26.18 \\ \hline
 900 & 1.95   & 5.97  & 170.1 & 0.829  & 1.84   & 35.82 \\ \hline
 1000 &2.45   & 7.47  & 220.3 & 1.05   & 2.34   & 46.82 \\ \hline
 & \multicolumn{6}{c|}{$Q^2 = m^2/4 \Rightarrow \mu_0^{\rm DIS} = 0.24m,
\mu_0^{\overline{\rm MS}} = 0.31m$} \\ \hline
 200 & 0.0099 & 0.016 & 0.201 & 0.0030 & 0.0041 & 0.046 \\ \hline
 300 & 0.106  & 0.200 & 2.35  & 0.035  & 0.056  & 0.571 \\ \hline
 400 & 0.369  & 0.735 & 8.33  & 0.132  & 0.218  & 2.10  \\ \hline
 500 & 0.813  & 1.66  & 18.63 & 0.306  & 0.516  & 4.80  \\ \hline
 600 & 1.41   & 2.94  & 32.98 & 0.557  & 0.948  & 8.63  \\ \hline
 700 & 2.14   & 4.51  & 50.25 & 0.868  & 1.49   & 13.36 \\ \hline
 800 & 2.97   & 6.33  & 70.60 & 1.24   & 2.13   & 19.03 \\ \hline
 900 & 3.85   & 8.26  & 92.73 & 1.65   & 2.85   & 25.16 \\ \hline
 1000 &4.78   & 10.32 & 116.5 & 2.10   & 3.63   & 32.12 \\ \hline
\end{tabular}
\caption[]{The hadronic $b \overline b$ production cross sections in
$pp$ interactions for $s_{\rm cut} = 1.2s_{4, {\rm min}}$, $m = 4.75$ 
GeV/$c^2$ and $\theta^* = 90^\circ$.}
\end{table}

\newpage
\begin{table}[htb]
\begin{tabular}{|c|c|c|c|c|c|c|} \hline
 & \multicolumn{3}{c|}{NLL} & \multicolumn{3}{c|}{LL} \\ \hline
$p_{\rm lab}$ (GeV/$c$) & $\sigma_{q \overline q}^{\rm DIS}$ (nb) & $\sigma_{q 
\overline q}^{\overline{\rm MS}}$ (nb) & $\sigma_{gg}$ (nb) & $\sigma_{q 
\overline q}^{\rm DIS}$ (nb) &  $\sigma_{q 
\overline q}^{\overline{\rm MS}}$ (nb) & $\sigma_{gg}$ (nb) \\ \hline
 & \multicolumn{6}{c|}{$Q^2 = m^2 \Rightarrow \mu_0^{\rm DIS} = 0.39m,
\mu_0^{\overline{\rm MS}} = 0.54m$} \\ \hline
 200 & 0.0035 & 0.0050 & 0.015 & 0.0017 & 0.0023 & 0.0071 \\ \hline
 300 & 0.037  & 0.061  & 0.187 & 0.019  & 0.029  & 0.094  \\ \hline
 400 & 0.131  & 0.225  & 0.723 & 0.072  & 0.111  & 0.373  \\ \hline
 500 & 0.295  & 0.517  & 1.72  & 0.166  & 0.262  & 0.910  \\ \hline
 600 & 0.525  & 0.932  & 3.20  & 0.303  & 0.482  & 1.70   \\ \hline
 700 & 0.811  & 1.46   & 5.14  & 0.478  & 0.765  & 2.76   \\ \hline
 800 & 1.14   & 2.06   & 7.46  & 0.685  & 1.10   & 4.05   \\ \hline
 900 & 1.51   & 2.74   & 10.18 & 0.918  & 1.48   & 5.54   \\ \hline
 1000 &1.92   & 3.48   & 13.20 & 1.17   & 1.90   & 7.24   \\ \hline
 & \multicolumn{6}{c|}{$Q^2 = 4m^2 \Rightarrow \mu_0^{\rm DIS} = 0.56m,
\mu_0^{\overline{\rm MS}} = 0.85m$} \\ \hline
 200 & 0.0028 & 0.0054 & 0.018 & 0.0013 & 0.0023 & 0.0079 \\ \hline
 300 & 0.029  & 0.059  & 0.216 & 0.015  & 0.027  & 0.099  \\ \hline
 400 & 0.102  & 0.209  & 0.825 & 0.054  & 0.097  & 0.388  \\ \hline
 500 & 0.228  & 0.471  & 1.99  & 0.126  & 0.227  & 0.945  \\ \hline
 600 & 0.402  & 0.837  & 3.72  & 0.227  & 0.409  & 1.78   \\ \hline
 700 & 0.620  & 1.29   & 6.02  & 0.356  & 0.644  & 2.93   \\ \hline
 800 & 0.874  & 1.83   & 8.85  & 0.509  & 0.920  & 4.31   \\ \hline
 900 & 1.16   & 2.43   & 12.17 & 0.684  & 1.23   & 5.93   \\ \hline
 1000 &1.46   & 3.06   & 15.94 & 0.873  & 1.58   & 7.88   \\ \hline
 & \multicolumn{6}{c|}{$Q^2 = m^2/4 \Rightarrow \mu_0^{\rm DIS} = 0.28m,
\mu_0^{\overline{\rm MS}} = 0.34m$} \\ \hline
 200 & 0.0025 & 0.0029 & 0.010 & 0.0013 & 0.0014 & 0.0057 \\ \hline
 300 & 0.033  & 0.045  & 0.159 & 0.018  & 0.023  & 0.090  \\ \hline
 400 & 0.133  & 0.188  & 0.646 & 0.076  & 0.101  & 0.376  \\ \hline
 500 & 0.315  & 0.461  & 1.57  & 0.188  & 0.255  & 0.940  \\ \hline
 600 & 0.580  & 0.866  & 2.93  & 0.356  & 0.489  & 1.78   \\ \hline
 700 & 0.915  & 1.38   & 4.69  & 0.576  & 0.801  & 2.89   \\ \hline
 800 & 1.31   & 1.99   & 6.81  & 0.837  & 1.17   & 4.25   \\ \hline
 900 & 1.76   & 2.70   & 9.26  & 1.14   & 1.60   & 5.85   \\ \hline
 1000 &2.24   & 3.46   & 11.92 & 1.47   & 2.08   & 7.58   \\ \hline
\end{tabular}
\caption[]{The hadronic $b \overline b$ production cross sections in
$pp$ interactions for $s_{\rm cut} = 1.6s_{4, {\rm min}}$, $m = 4.75$ 
GeV/$c^2$ and  $\theta^* = 90^\circ$.}
\end{table}

\newpage
\begin{center}
{\bf Figure Captions}
\end{center}
\vspace{0.2in}

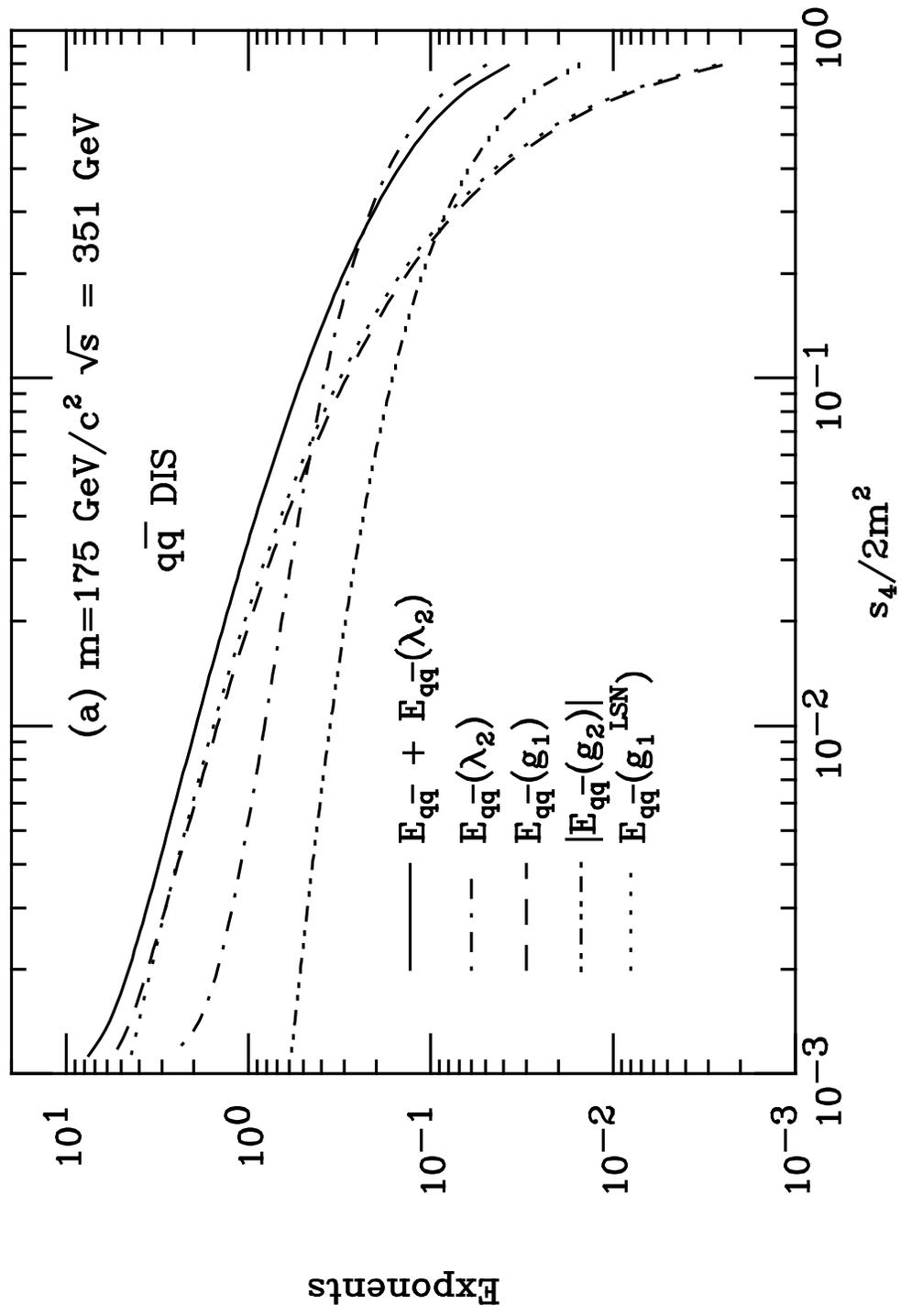
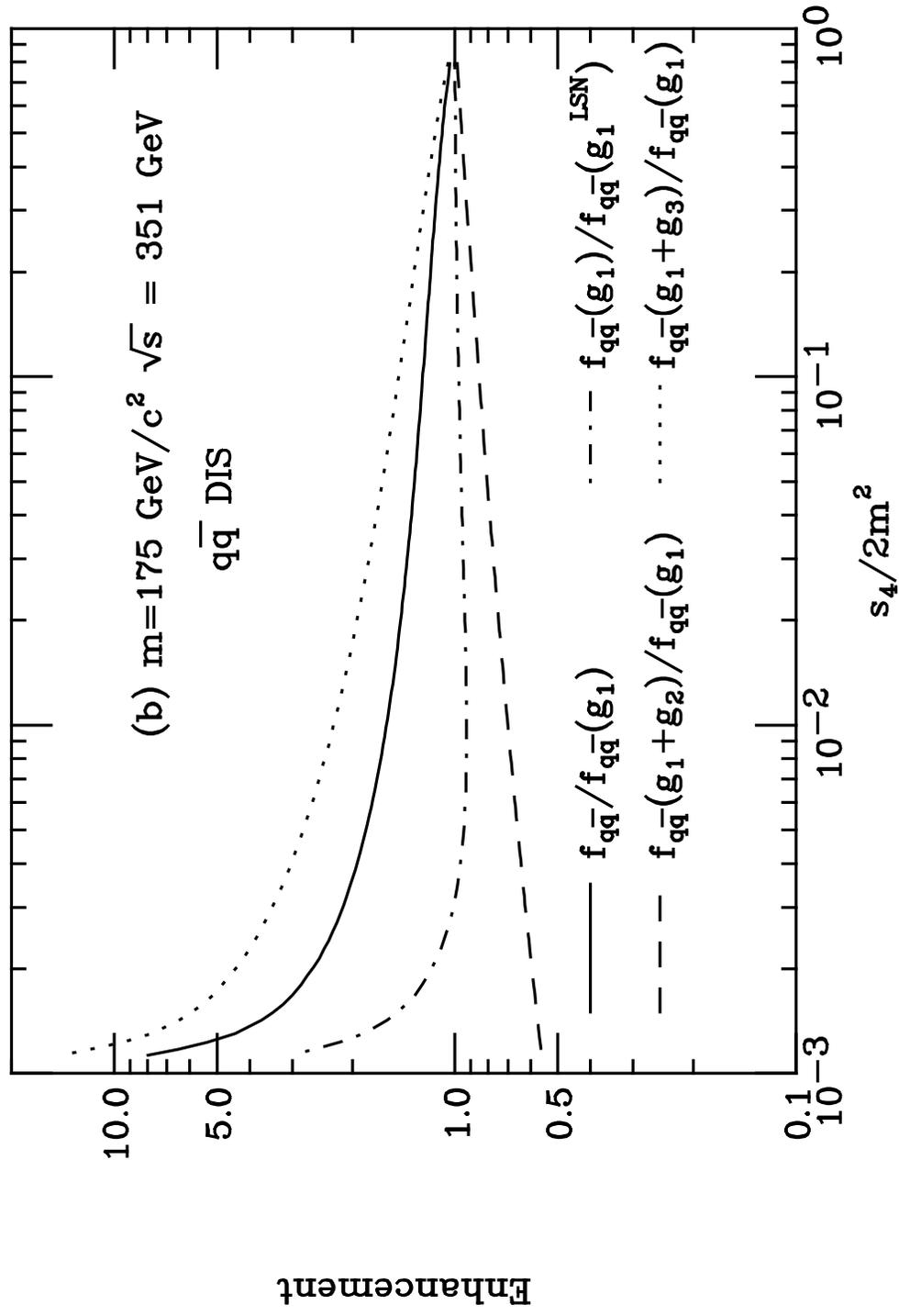
\noindent Figure 1.  (a) We show the contributions to $f_{q \overline q}^{\rm 
DIS}$ in top quark production for $m= 175$ GeV/$c^2$ and $\sqrt{s} = 351$
GeV as a function of $s_4/2m^2$.  The solid curve
shows $\ln(f_{q \overline q}^{\rm DIS})$ for the octet component at $90^\circ$,
the dot-dashed curve, $E_{q \overline q}(\lambda_2)$, the dashed curve,
$E_{q \overline q}^{\rm DIS}(g_1)$, and the dot-dot-dashed curve,
$|E_{q \overline q}^{\rm DIS}(g_2)|$. The dotted curve shows 
$E^{\rm DIS}$ as defined in eq.\ (3.30) of Ref.\ \cite{LSN1}.
(b) The ratios of the exponential factors are given to show the
enhancement relative to $f_{q \overline q}(g_1)$ where $f_{q
\overline q}(g_1) \equiv \exp(E_{q \overline 
q}^{\rm DIS}(g_1))$.  The solid curve shows $f_{q \overline q}^{\rm DIS}/f_{q 
\overline q}(g_1)$, the dashed curve, $f_{q \overline q}(g_1 + 
g_2)/f_{q \overline q}(g_1)$, and the dotted curve, $f_{q \overline q}(g_1 + 
g_3(\lambda_2))/f_{q \overline q}(g_1)$.  The ratio $f_{q \overline q}(g_1)/
\exp(E^{\rm DIS})$, given in the dot-dashed curve, shows the enhancement
of $g_1$ from eq.\ (2.5) relative to the one-loop value in \cite{LSN1}.\\[2ex]

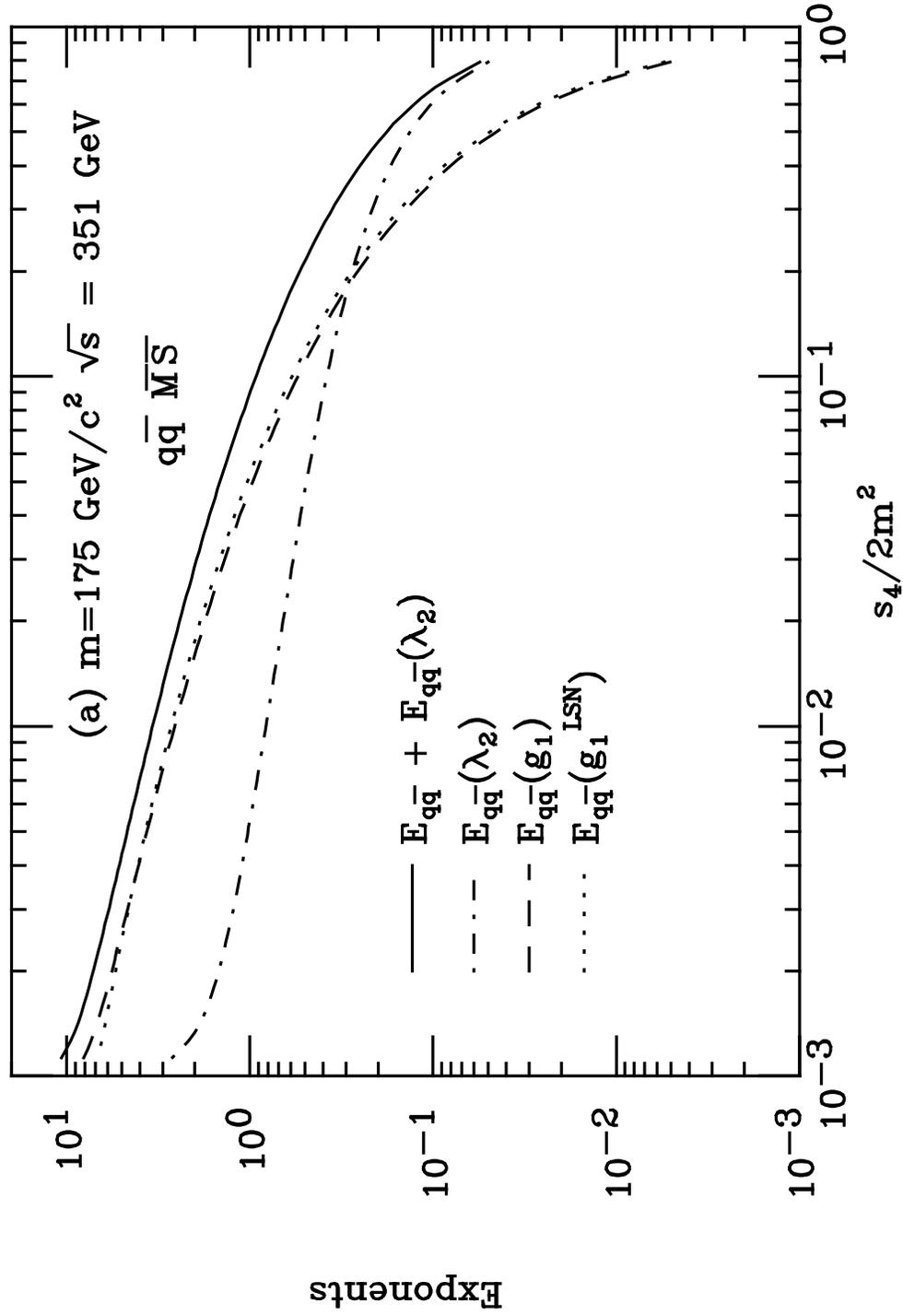
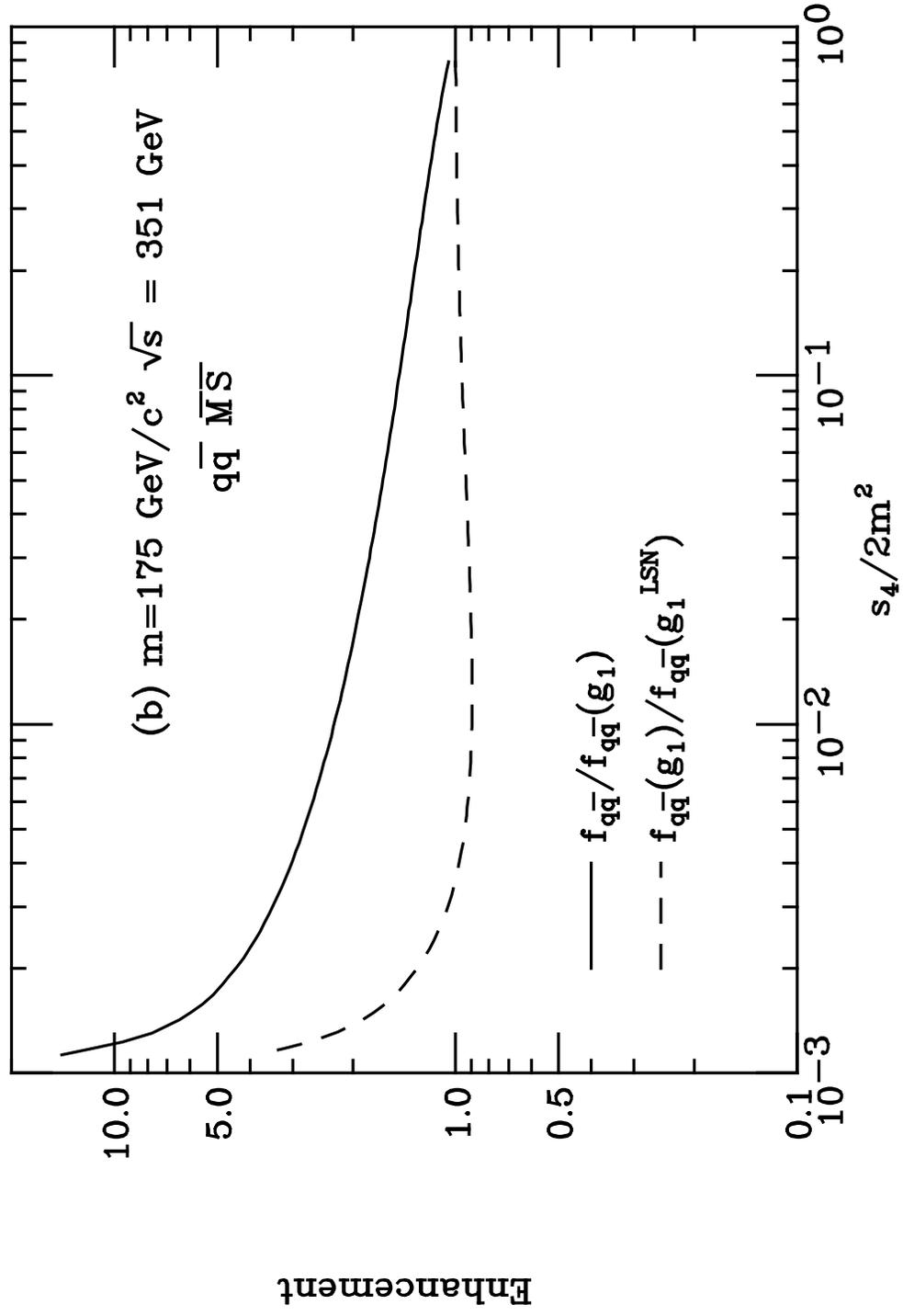
\noindent Figure 2. (a) We show the contributions to 
$f_{q \overline q}^{\overline
{\rm MS}}$ in top quark production for $m=175$ GeV/$c^2$ and $\sqrt{s} = 351$
GeV as a function of $s_4/2m^2$.  The solid curve
shows $\ln(f_{q \overline q}^{\overline {\rm MS}})$ 
for the octet component at $90^\circ$,
the dot-dashed curve, $E_{q \overline q}(\lambda_2)$, and the dashed curve,
$E_{q \overline q}^{\overline {\rm MS}}(g_1)$.  The dotted curve shows 
$E^{\overline {\rm MS}}$ as defined in eq.\ (3.34) of Ref.\ \cite{LSN1}.
(b) The ratios of the exponential factors are given to show the
enhancements.  The solid curve shows $f_{q \overline q}^{\overline {\rm 
MS}}/f_{q \overline q}(g_1)$ while the dashed curve is the ratio $f_{q 
\overline q}(g_1)/\exp(E^{\overline {\rm MS}})$ where $f_{q \overline q}(g_1)
\equiv \exp(E_{q \overline q}^{\overline {\rm MS}}(g_1))$.\\[2ex]

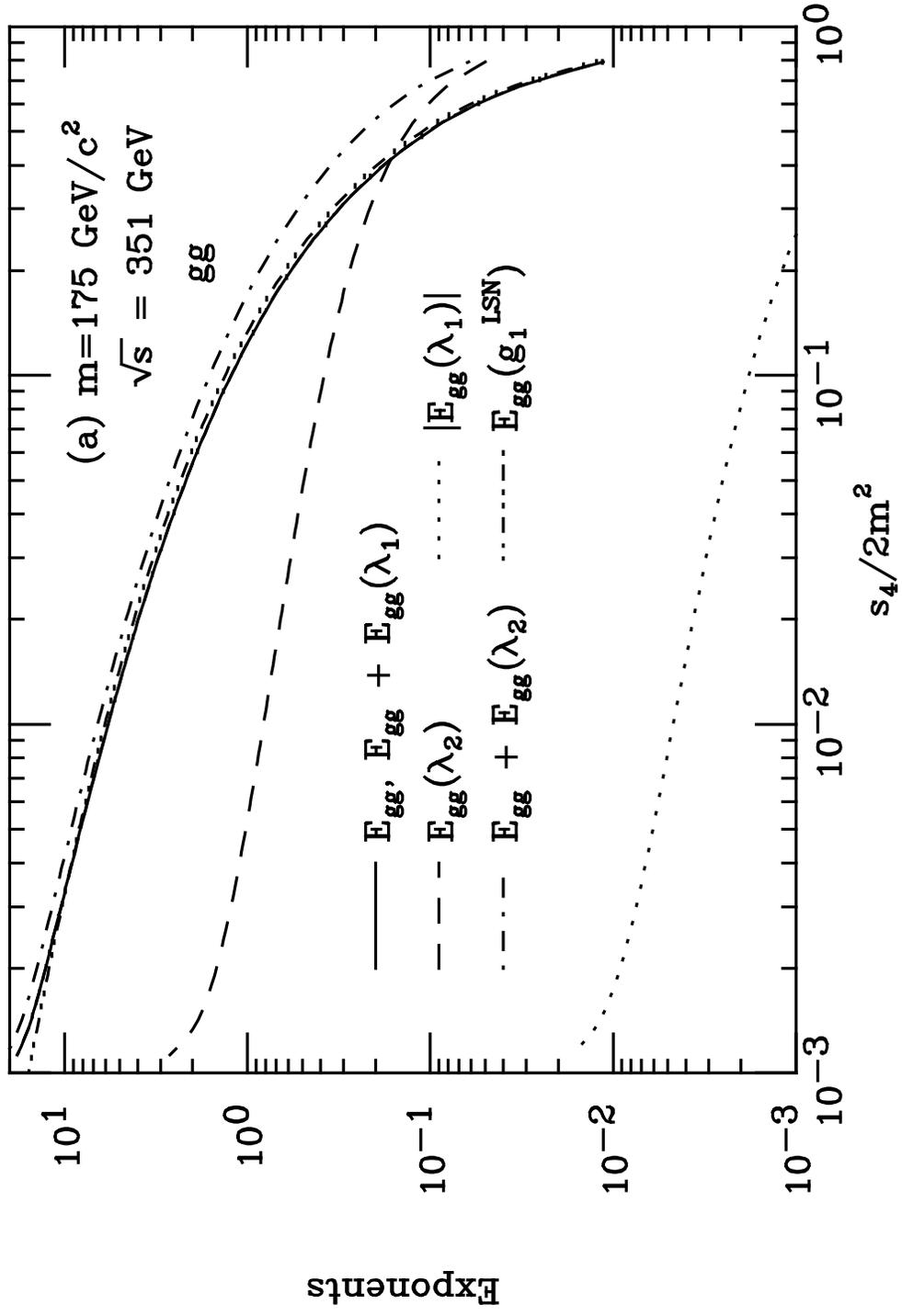
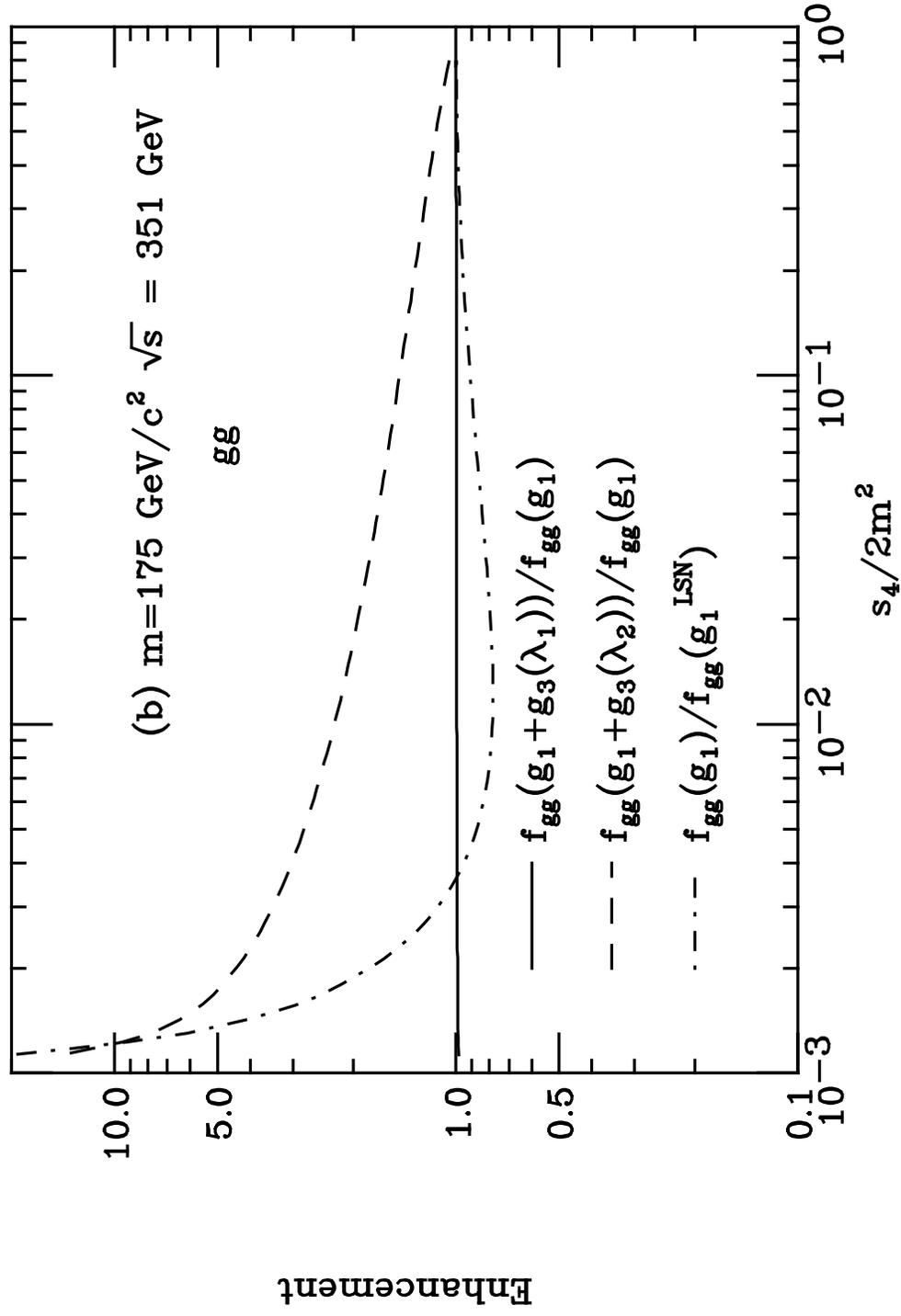
\noindent Figure 3. (a) We show the contributions to $f_{gg}$
in top quark production for $m= 175$ GeV/$c^2$ 
and $\sqrt{s} = 351$ GeV as a function of $s_4/2m^2$.  The solid curve
shows $E_{gg}$; the dashed and dotted curves give the contributions from the
eigenvalues, $E_{gg}(\lambda_2)$ and $|E_{gg}(\lambda_1)|$ respectively.
The dot-dashed curve shows the sum $E_{gg}+E_{gg}(\lambda_2)$; the sum $E_{gg}
+E_{gg}(\lambda_1)$ is indistinguishable from the solid curve.
The dot-dot-dashed curve shows 
$E^{\overline {\rm MS}}$ for the gluon from Ref.\ \cite{LSN1}.
(b) The ratios of the exponential factors are given to show the
enhancements relative to $f_{gg}(g_1) \equiv \exp(E_{gg})$.  The solid curve
shows $f_{gg}(g_1 + g_3(\lambda_1))/f_{gg}(g_1)$ and the dashed curve, 
$f_{gg}(g_1 + g_3(\lambda_2))/f_{gg}(g_1)$.  The ratio of $f_{gg}(g_1)$
to $\exp(E^{\overline {\rm MS}})$, given in the dot-dashed curve, compares
the $g_1$ from eq.\ (2.5) relative to the one-loop value in \cite{LSN1}.\\[2ex]

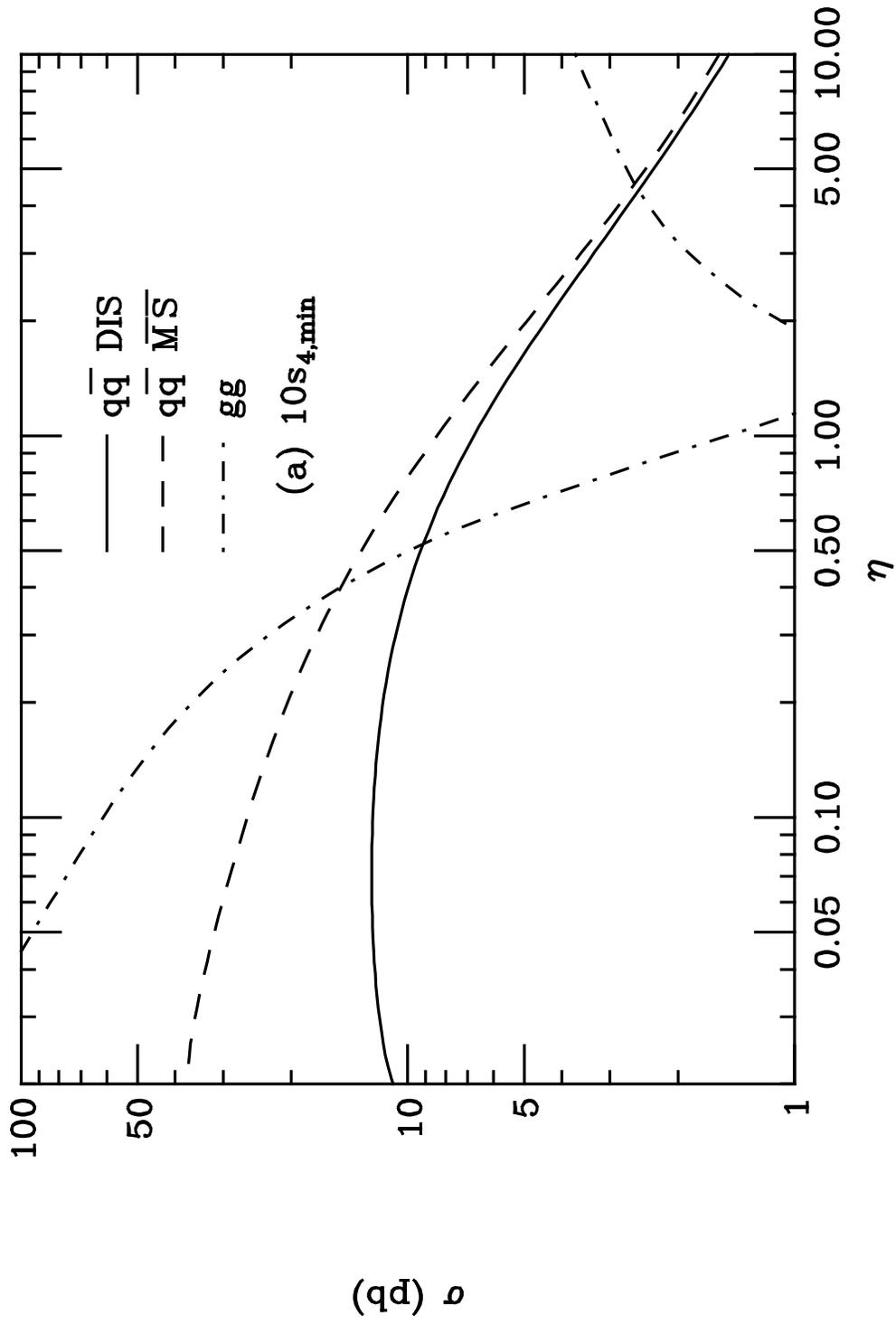
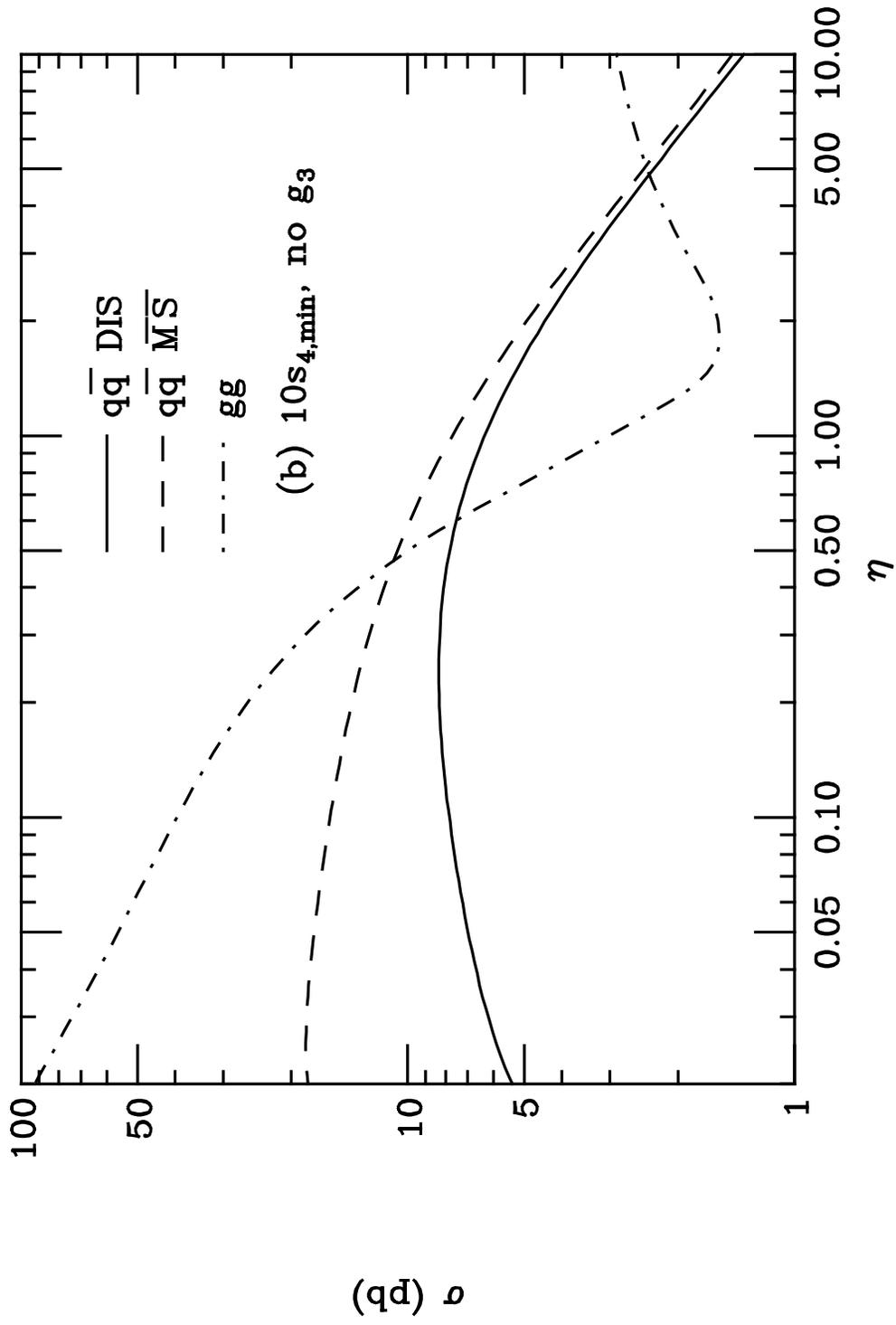
\noindent Figure 4.  The partonic top quark production cross section is
shown as a function of $\eta= (s- 4m^2)/(4m^2)$ 
for the $q \overline q$ channel in the DIS
(solid) and $\overline{\rm MS}$ (dashed) schemes and the $gg$ (dot-dashed)
channel. We use $s_{\rm cut} \approx 10s_{4, {\rm 
min}}$ with (a) and without (b) 
the NLL $g_3$ contributions respectively.\\[2ex]

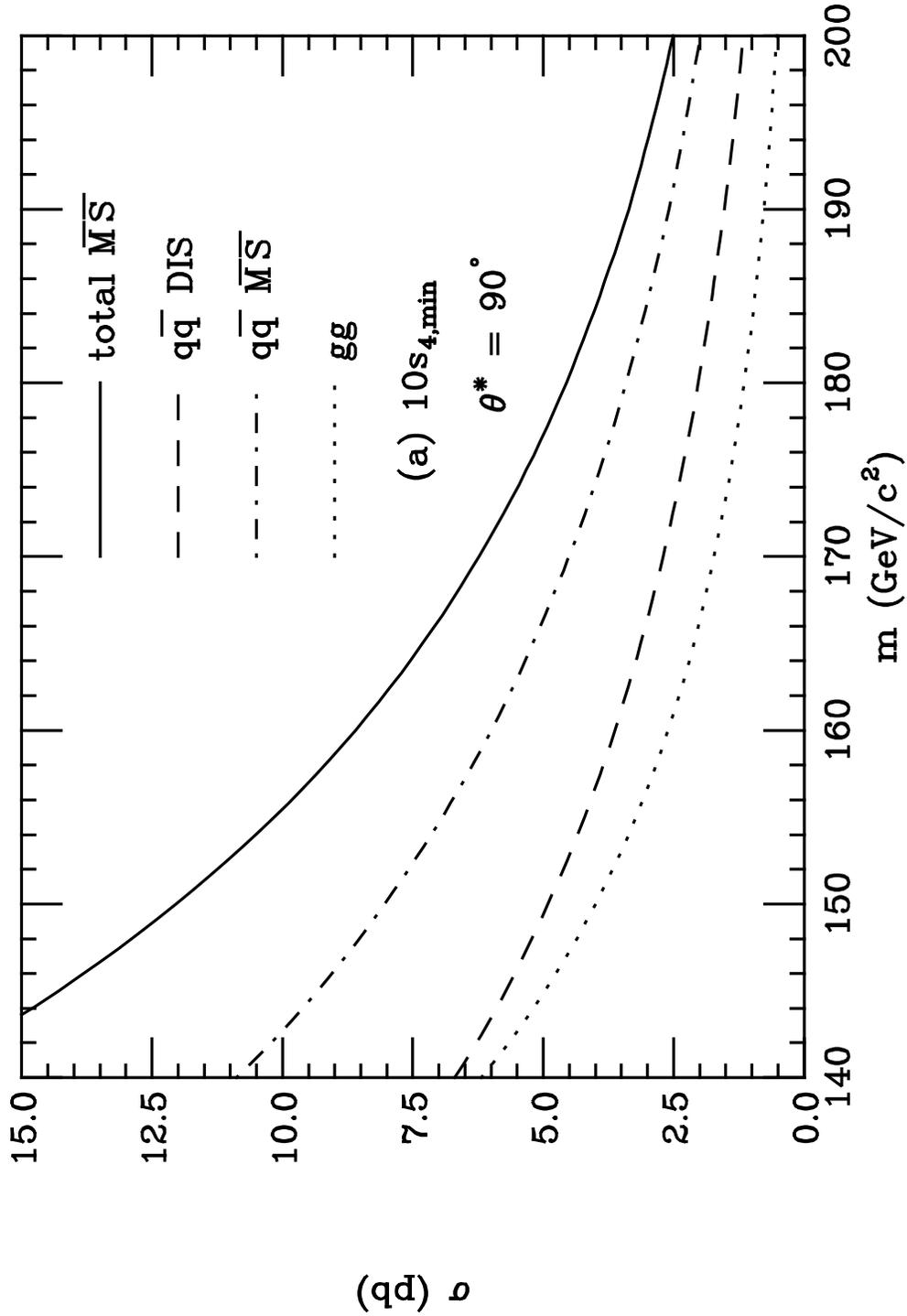
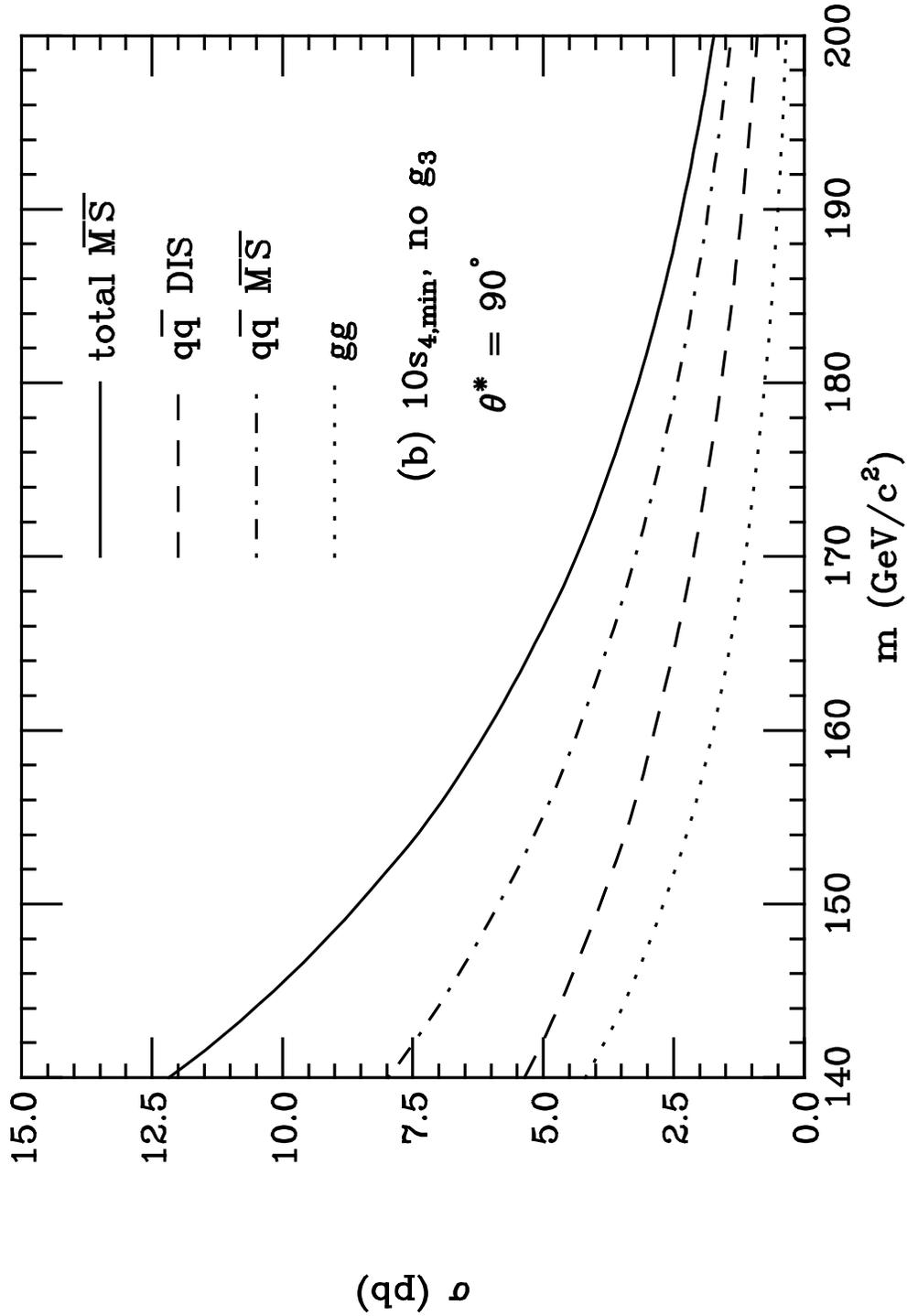
\noindent Figure 5.  The hadronic $t \overline t$ production cross section 
at $\theta^* = 90^\circ$ is
given as a function of top quark mass for $p \overline p$ collisions at the
Tevatron energy, $\sqrt{S} = 1.8$ TeV.  We use the MRS D$-^\prime$ parton
densities in the DIS scheme for $q \overline q$ annihilation (dashed)
and in the $\overline{\rm MS}$ scheme for the $q \overline q$ (dot-dashed)
and $gg$ (dotted) channels. 
The sum of the $q \overline q$ and $gg$ channels in the $\overline{\rm MS}$
scheme is given in the solid curve. The results are given for
$s_{\rm cut} \approx 10 s_{4, {\rm min}}$ with the NLL $g_3$
contributions in (a) and with these excluded in (b).\\[2ex]

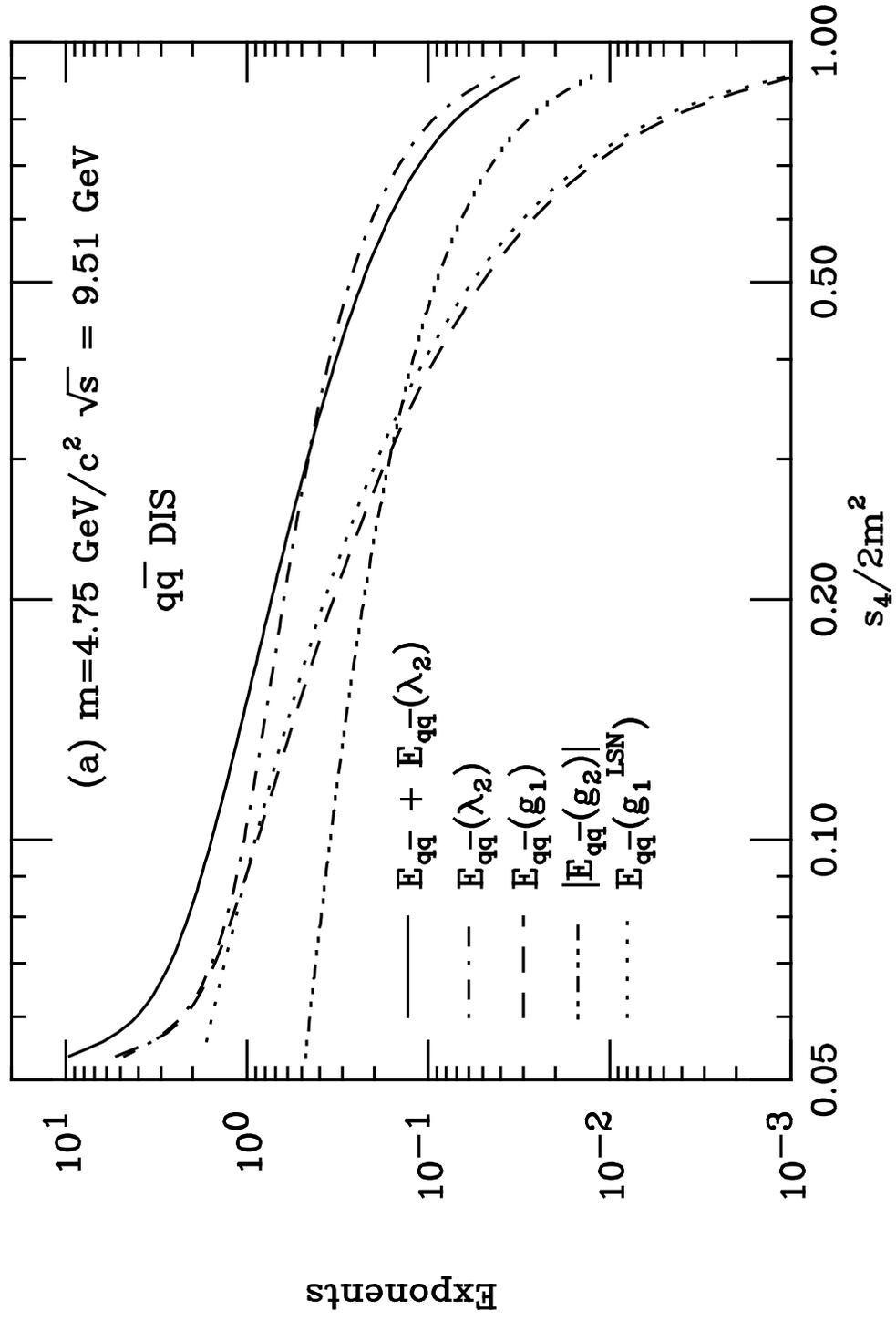
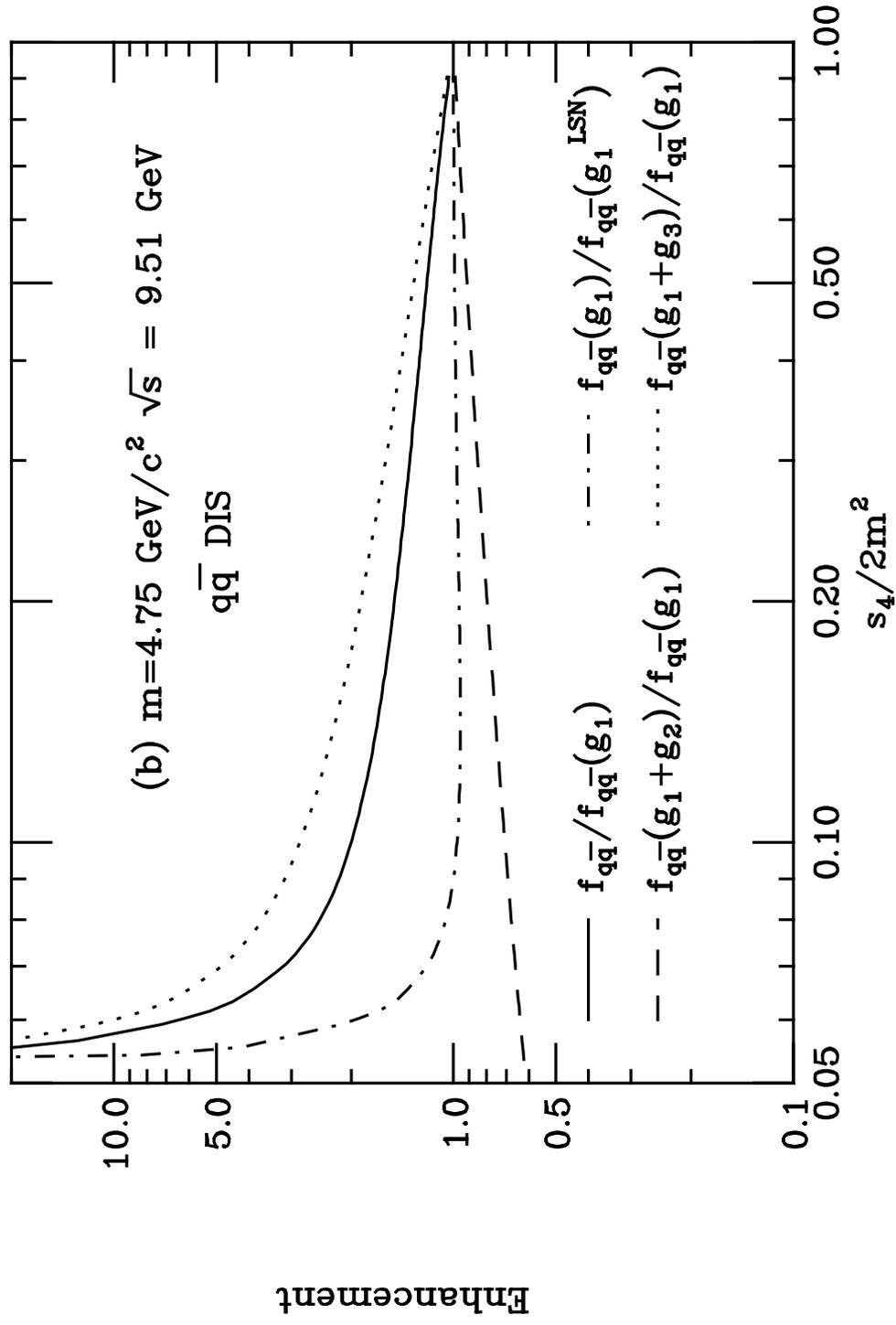
\noindent Figure 6.  We show the contributions to $f_{q \overline q}^{\rm 
DIS}$ and the enhancement factors in bottom quark production for
$m= 4.75$ GeV/$c^2$ and $\sqrt{s} = 9.51$
GeV.  The labels are as in fig.\ 1.\\[2ex]

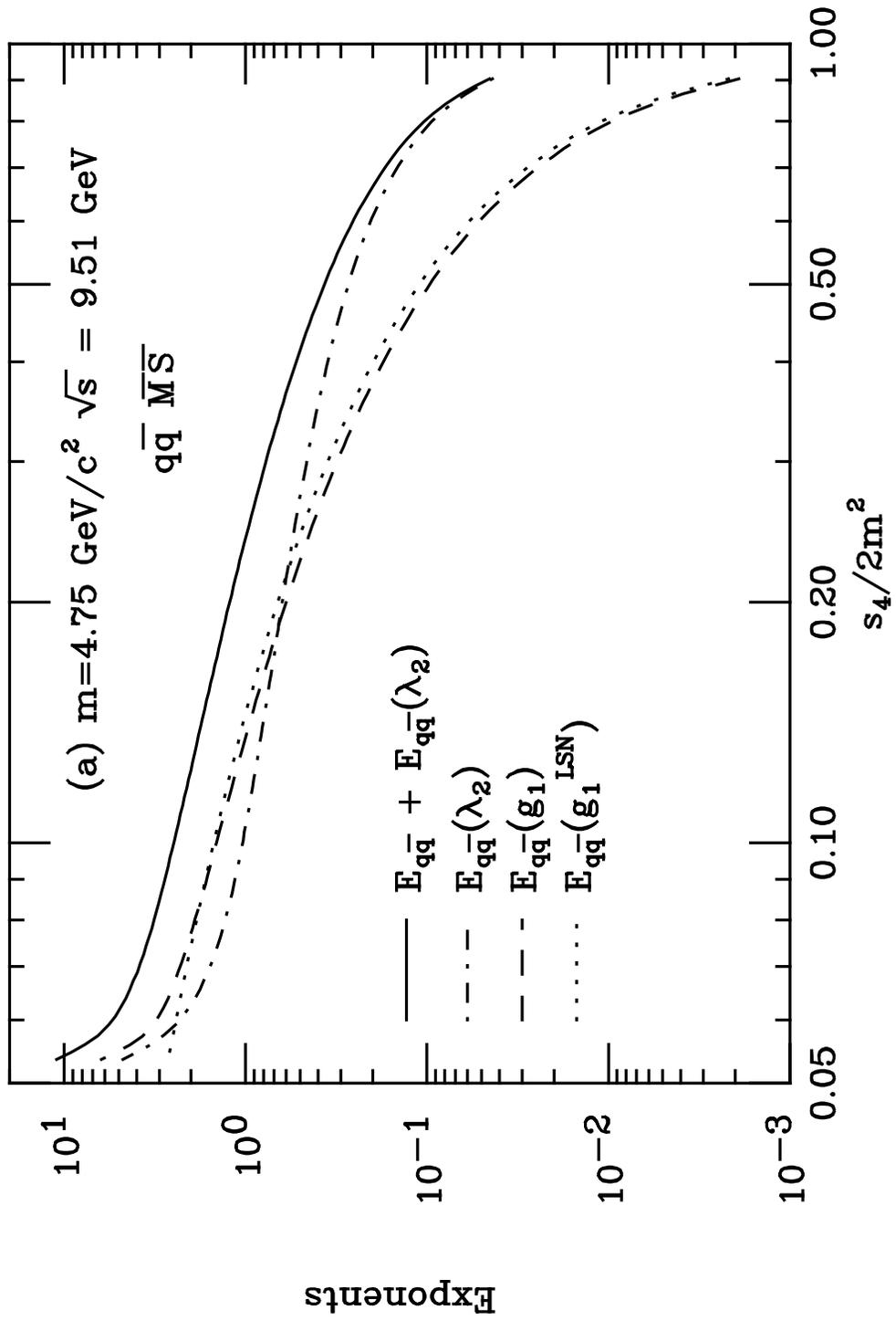
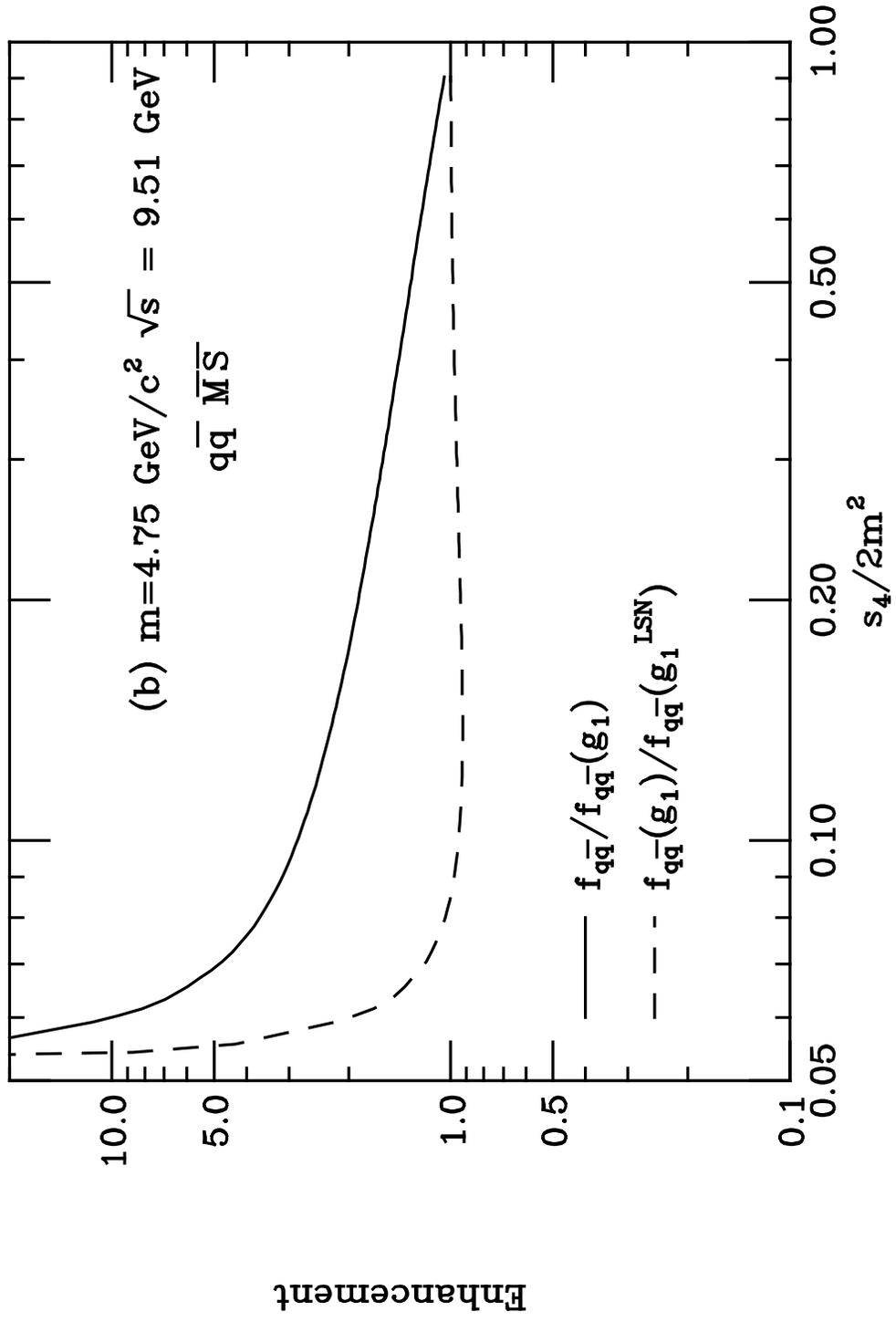
\noindent Figure 7.   We show the contributions to $f_{q \overline
q}^{\overline {\rm MS}}$ and the enhancement factors 
in bottom quark production for $m= 4.75$ GeV/$c^2$ and 
$\sqrt{s} = 9.51$ GeV.  The labels are as in fig.\ 2.\\[2ex]

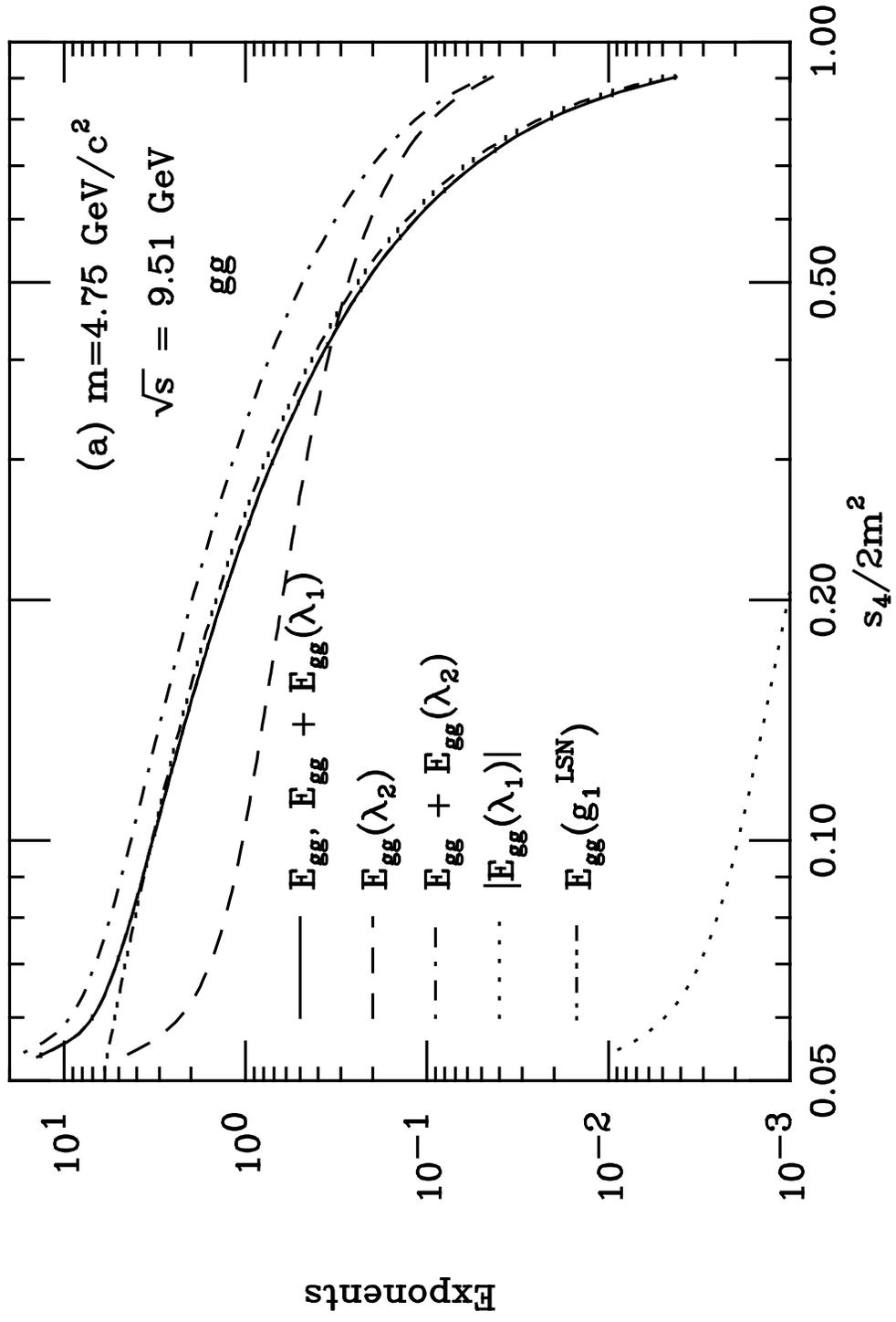
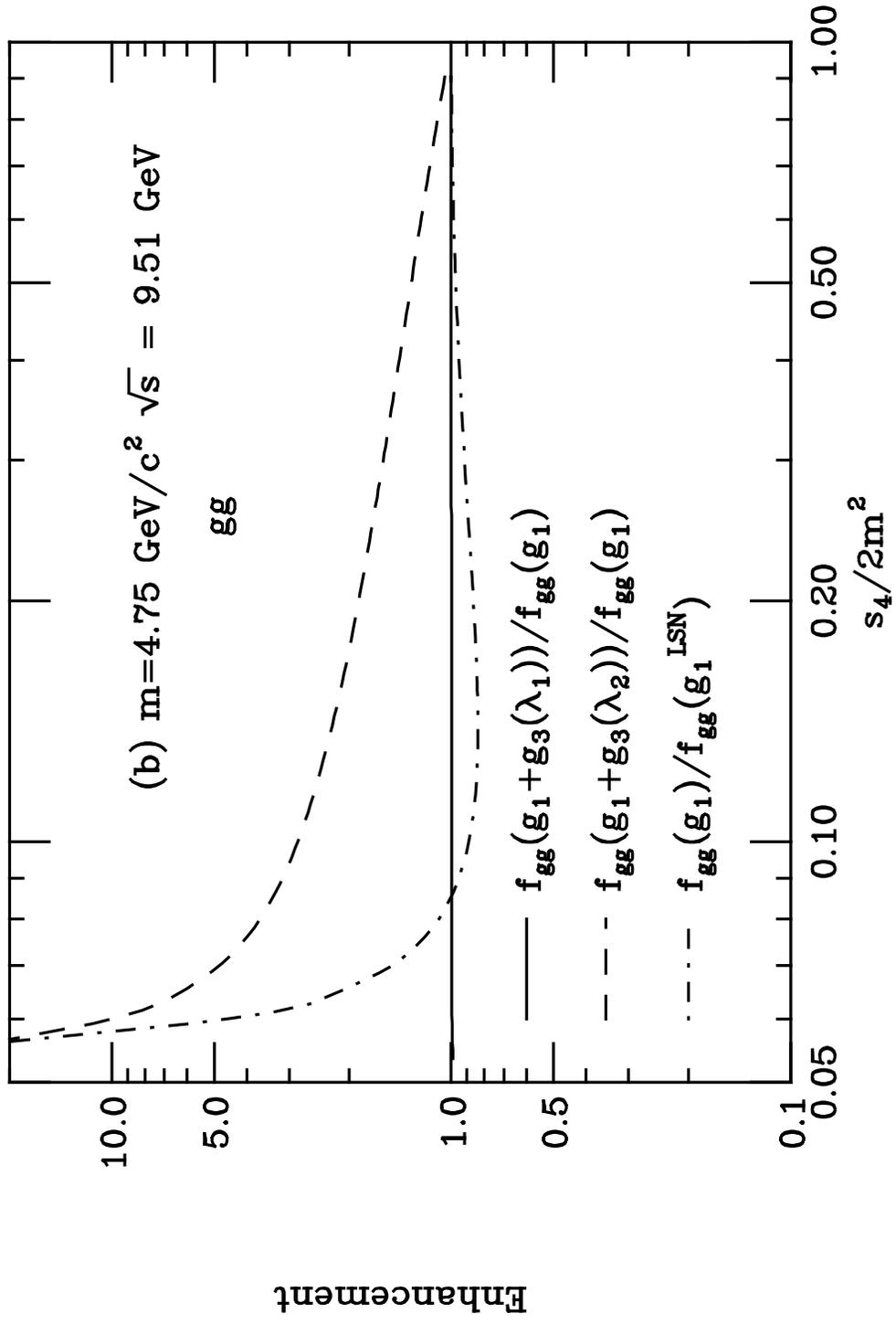
\noindent Figure 8.   We show the contributions to $f_{gg}$ and the enhancement
factors in bottom quark production for
$m= 4.75$ GeV/$c^2$ and 
$\sqrt{s} = 9.51$ GeV.  The labels are as in fig.\ 3.\\[2ex]

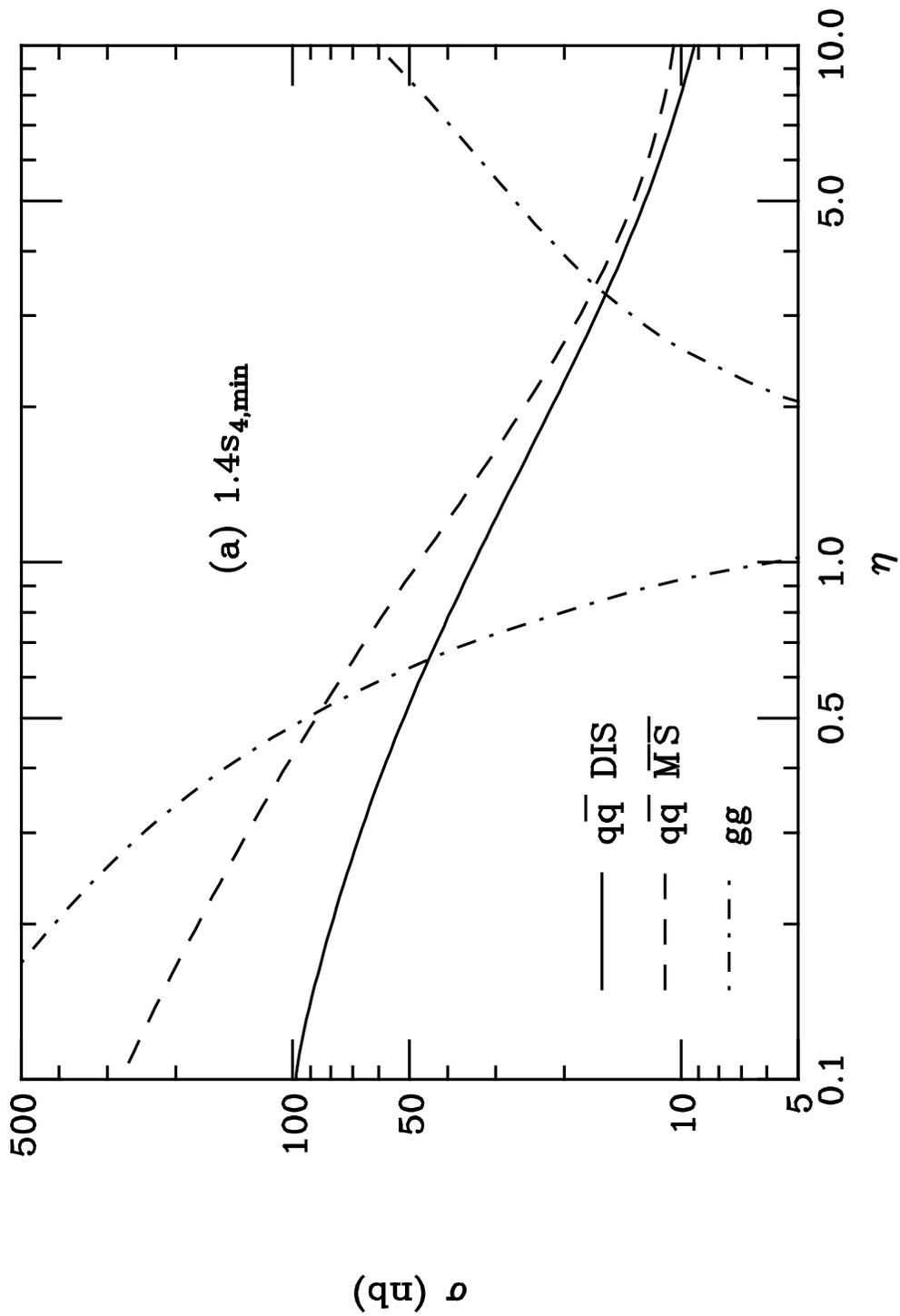
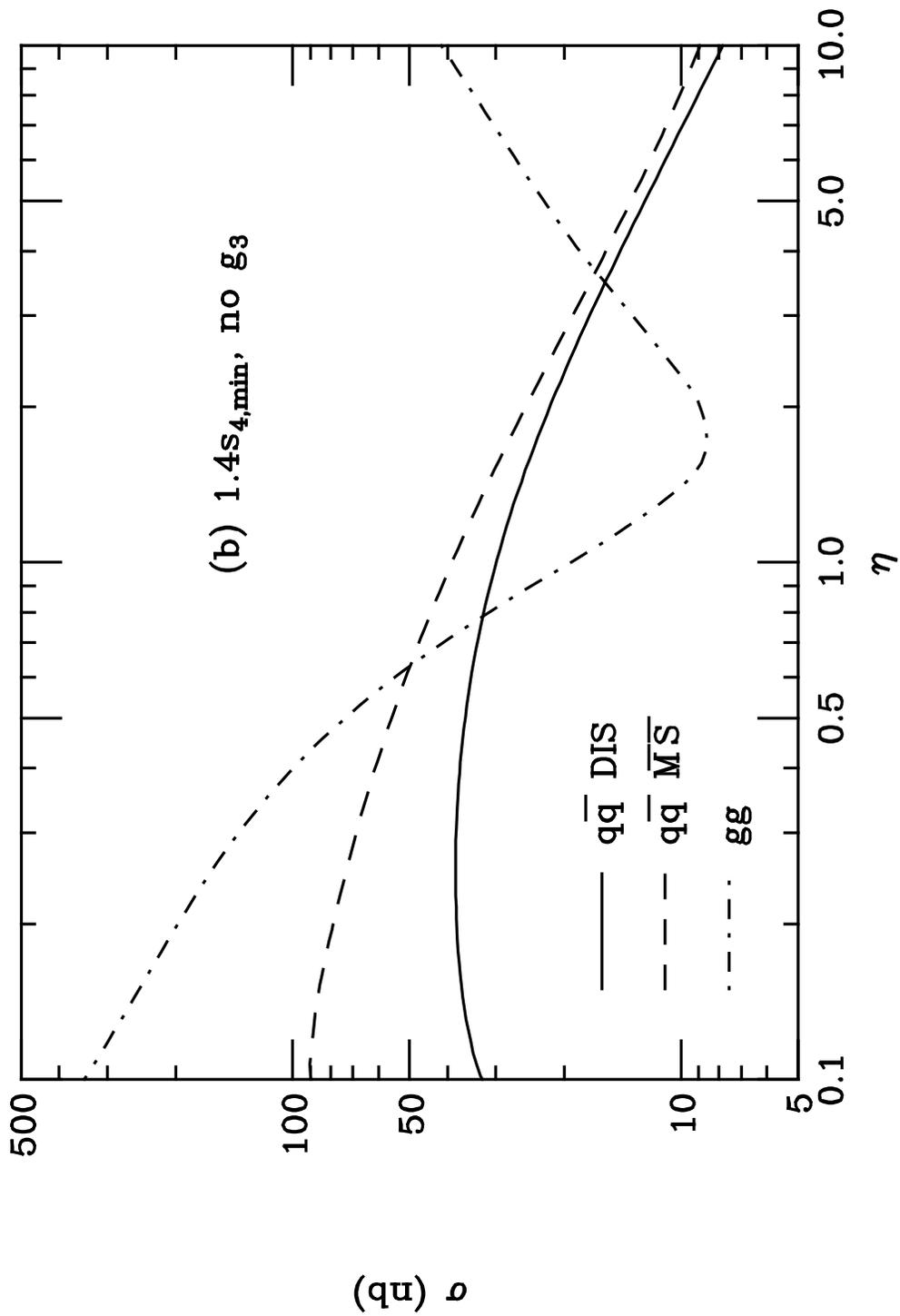
\noindent Figure 9.  The partonic bottom quark production cross section is
shown as a function of $\eta$.  The labels are as in fig.\ 4.
We use $s_{\rm cut} \approx 1.4s_{4, {\rm min}}$ with the NLL
$g_3$ contributions in (a) and without them in (b).\\[2ex]

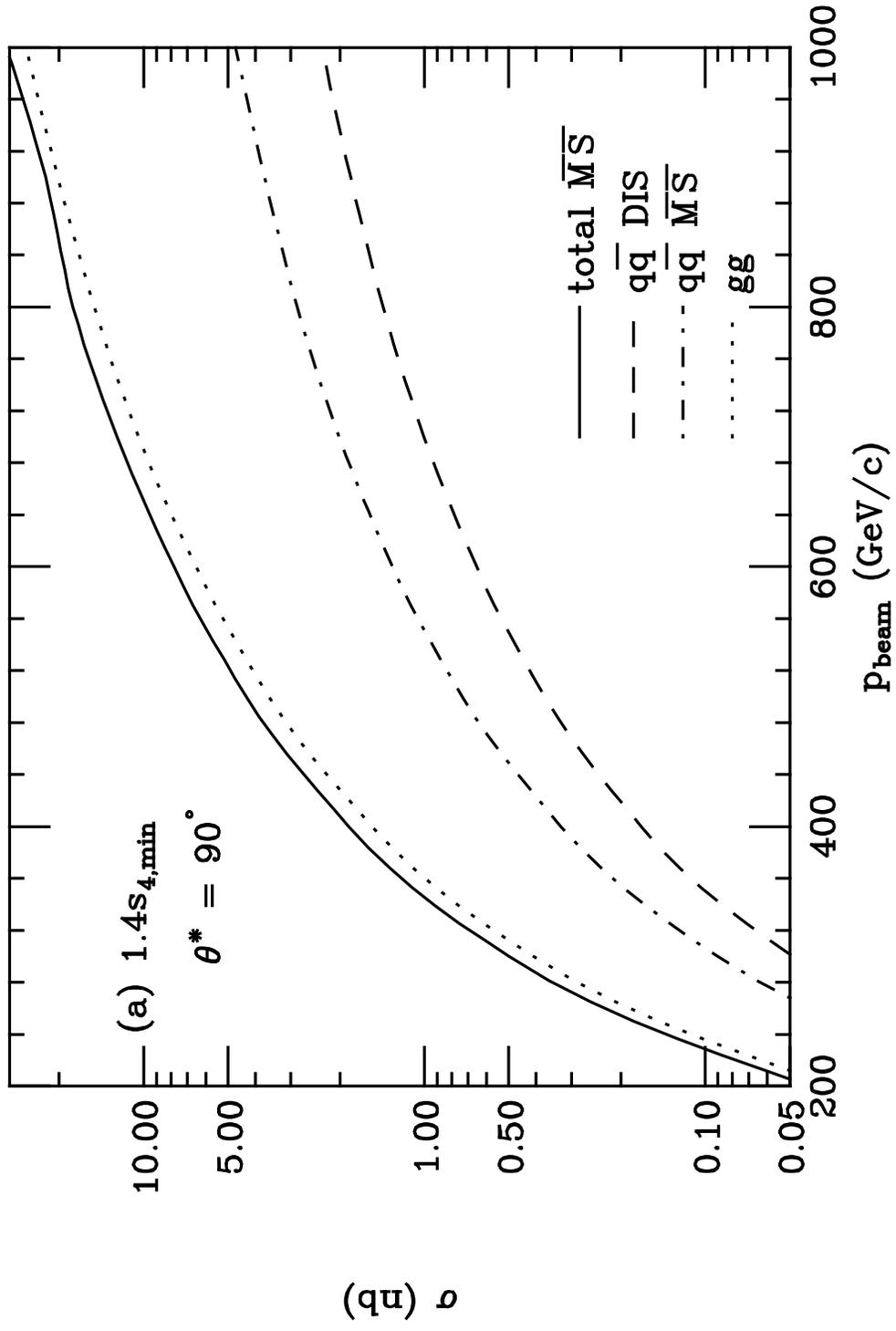
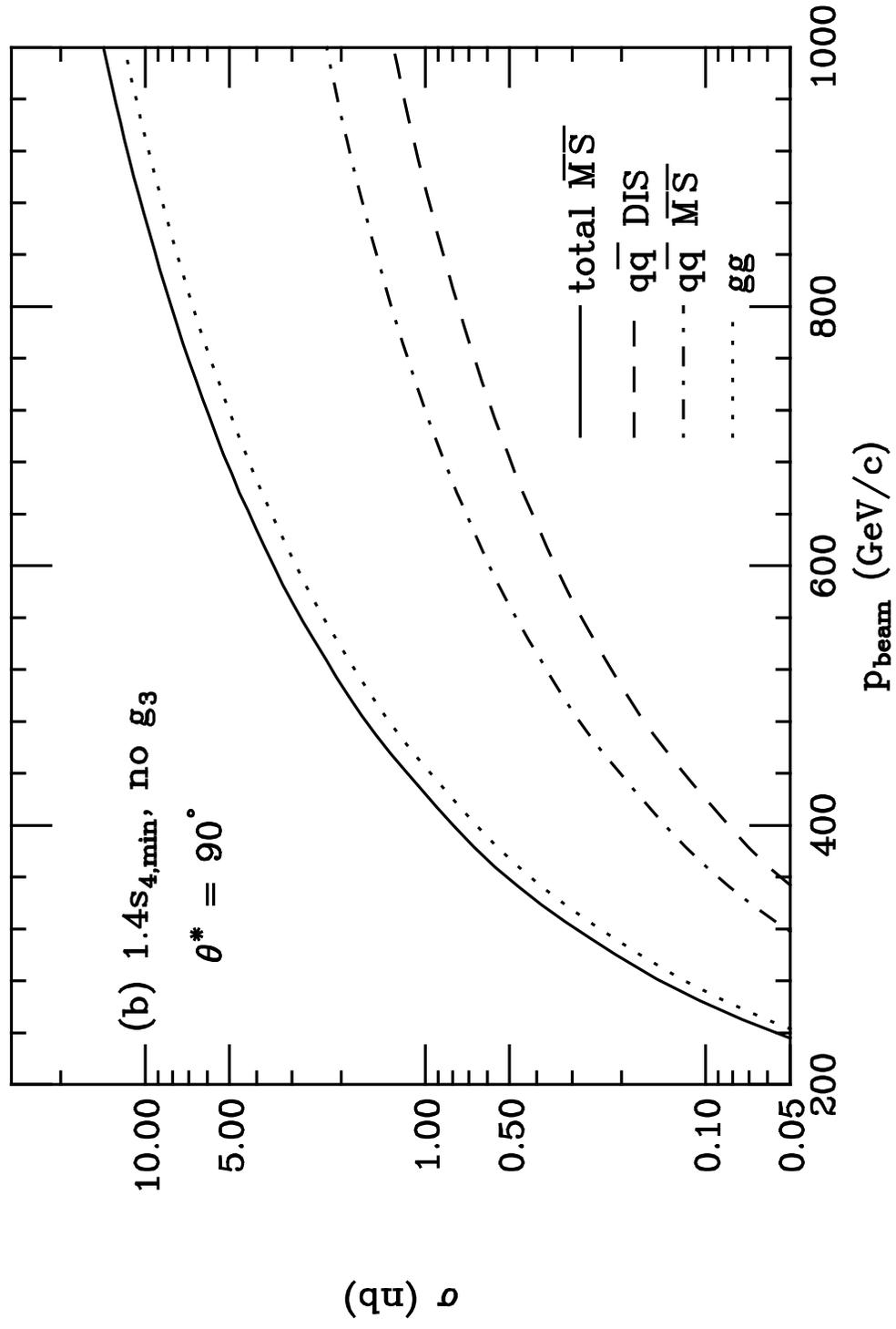
\noindent Figure 10.  The hadronic $b \overline b$ production cross section at
$\theta^* = 90^\circ$ is
given as a function of beam momentum for $pp$ interactions.    
We use the MRS D$-^\prime$ parton
densities in the DIS scheme for $q \overline q$ annihilation (dashed)
and in the $\overline{\rm MS}$ scheme for the $q \overline q$ (dot-dashed) 
and $gg$ (dotted) channels. The $\overline{\rm MS}$ sum is given in the 
solid curve.  The results are given for $s_{\rm cut} \approx
1.4s_{4, {\rm min}}$ with the NLL $g_3$ corrections in (a) and without them in 
(b).

\end{document}